\documentclass[12pt,a4paper]{article}

\usepackage{amssymb}
\usepackage{epsf}
\begin{document}

\title{Hydrodynamics of ultra-relativistic bubble walls}
\author{\large Leonardo Leitao\thanks{%
E-mail address: lleitao@mdp.edu.ar}~  and Ariel M\'{e}gevand\thanks{%
Member of CONICET, Argentina. E-mail address: megevand@mdp.edu.ar} \\[0.5cm]
\normalsize \it IFIMAR (CONICET-UNMdP)\\
\normalsize \it Departamento de F\'{\i}sica, Facultad de Ciencias Exactas
y Naturales, \\
\normalsize \it UNMdP, De\'{a}n Funes 3350, (7600) Mar del Plata, Argentina
}
\date{}
\maketitle

\begin{abstract}
In cosmological first-order phase transitions, gravitational waves are
generated by the collisions of bubble walls and by the bulk motions caused in
the fluid. A sizeable signal may result from fast-moving walls. In this work we
study the hydrodynamics associated to the fastest propagation modes, namely,
ultra-relativistic detonations and runaway solutions. We compute the energy
injected by the phase transition into the fluid and the energy which
accumulates in the bubble walls. We provide analytic approximations and fits as
functions of the net force acting on the wall, which can be readily evaluated
for specific models. We also study the back-reaction of hydrodynamics on the
wall motion, and we discuss  the extrapolation of the friction force away
from the ultra-relativistic limit. We use these results to estimate the
gravitational wave signal from detonations and runaway walls.
\end{abstract}

\section{Introduction}

A first-order phase transition of the universe proceeds by nucleation and
expansion of bubbles, and may have different cosmological consequences,
depending on the velocity of bubble growth. For instance, the generation of the
baryon asymmetry of the universe in the electroweak phase transition is most
efficient for non-relativistic bubble walls, and is suppressed as the bubble
wall velocity approaches the speed of sound in the plasma \cite{baryo}. In
contrast, the formation of gravitational waves may be sizeable if the wall
velocity is supersonic \cite{gw}. These cosmological consequences generally
depend not only on the wall velocity but also on the bulk motions of the plasma
caused by the wall. For instance, gravitational waves are generated by bubble
collisions \cite{gw,gwcol,hk08} as well as by turbulence
\cite{gwturb,cds09,cds10} and sound waves \cite{gwsound}.

The propagation of the phase transition fronts (bubble walls) is affected by
hydrodynamics in a non-trivial manner (see, e.g.,
\cite{hidrowall,ikkl94,ms09,kn11,inst}). The wall motion is driven essentially
by the difference of pressure between the two phases. This force grows with the
amount of supercooling, i.e., the further down the temperature descends below
the critical temperature, the larger the pressure difference between phases. As
a consequence, the driving force is very sensitive to the (inhomogeneous)
reheating which occurs due to the release of latent heat.

Besides, the microscopic interactions of the particles of the plasma with the
wall cause a friction force on the latter (see, e.g., \cite{mp95}). Computing
the friction force is a difficult task, and for many years only the
non-relativistic (NR) case was studied \cite{nrfric}. In this approximation, a
wall velocity $v_{w}\ll 1$ is assumed, and the friction force scales as
$v_{w}$. Beyond the NR regime, a dependence $v_w\gamma_w$ was usually assumed,
where $\gamma_w=1/\sqrt{1-v_{w}^{2}}$. As a consequence of this scaling, the
wall would always reach a terminal velocity. More recently, the \emph{total}
force acting on the wall was calculated in the ultra-relativistic (UR) limit,
$\gamma _{w}\gg 1$ \cite{bm09}. The result does not allow to discriminate the
friction or the hydrodynamic effects. Nevertheless, the net force
$F_{\mathrm{net}}$ is independent of $v_{w}$, which means that the friction
saturates as a function of $v_w\gamma _{w}$. As a consequence, the wall may run
away. For intermediate velocities, microscopic calculations of the friction
were hardly attempted \cite{hs13,knr14}. To compute the wall velocity,
phenomenological interpolations between the NR and the UR limits have been
considered in Refs.~\cite{ekns10,ariel13}.

Leaving aside the determination of the wall velocity, the perturbations caused
in the plasma by the moving wall have been extensively studied for the case of
a stationary solution \cite{kurki85,hidro,lm11,lm15}. Different hydrodynamic
regimes can be established, depending on  the wall velocity. For a subsonic
wall the hydrodynamic solution is a weak deflagration, in which the wall is
preceded by a shock wave. For a supersonic wall, we have a Jouguet deflagration
if the wall velocity is smaller than the Jouguet velocity. In this case, the
fluid is disturbed both in front and behind the wall. For higher wall
velocities, the solution is a weak detonation. For the detonation, the velocity
is so high that the fluid in front of the wall is unaffected. In this case, the
wall is followed by a rarefaction wave.

The steady-state hydrodynamics can be investigated as a function of
thermodynamic parameters (such as the latent heat) and of the wall velocity
(i.e., considering $v_w$ as a free parameter). Thus, in particular, the kinetic
energy in bulk motions of the plasma, which is relevant for the generation of
gravitational waves, was computed in Refs.~\cite{ekns10,lm11} for the whole
velocity range $0<v_w<1$. These results are useful for applications, as they do
not depend on a particular calculation of the wall velocity for a specific
model.

For the runaway case, the hydrodynamics was considered in Ref.~\cite{ekns10}.
However, the results rely on the decomposition of the total force into driving
and friction forces, and are sensitive to approximations. The decomposition of
the UR force was discussed also in Ref.~\cite{ariel13}. Since the net force is
known \cite{bm09}, it is actually not necessary, in the UR limit, to determine
the friction component in order to study the wall motion. However, identifying
the forces acting on the wall is useful, in the first place, to understand the
hydrodynamics, and, in the second place, to construct a phenomenological model
for the friction, which allows to compute the wall velocity away from the UR
limit.

In this paper we consider ultra-relativistic walls and we study, on the one
hand, the hydrodynamics as a function of the wall acceleration, and, on the
other hand, the role of hydrodynamics and friction in the determination of the
net force. In particular, we obtain the energy in bulk fluid motions as a
function of the net force $F_{\mathrm{net}}$, in the whole range of runaway
solutions. We also discuss the effect of reheating on the force, and we compare
with approximations used in previous approaches. We apply these results to the
estimation of the gravitational wave signal from phase transitions.

The paper is organized as follows. In Sec.~\ref{review} we review the dynamics
of the Higgs-fluid system. In Sec.~\ref{hidro} we consider the hydrodynamics of
detonations and runaway walls for given values of the wall velocity and
acceleration, while in Sec.~\ref{micro} we consider the wall equation of motion
and we analyze the dependence of the energy distribution on thermodynamic and
friction parameters. In Sec.~\ref{gw} we estimate the amplitude of the
gravitational waves as a function of all these quantities. We summarize our
conclusions in Sec.~\ref{conclu}. In appendix \ref{apend} we find analytic
results for the efficiency factor for the case of planar walls, and we provide
fits for the case of spherical walls.

\section{The Higgs-fluid system \label{review}}

To describe the phase transition, we shall consider a system consisting of an
order-parameter field $\phi (x)$ (the Higgs field) and a relativistic fluid
(the hot plasma). The latter is characterized by a four-velocity field $ u^{\mu
}(x)$ and the temperature $T(x)$. The phase transition dynamics is mostly
determined by the free-energy density, also called finite-temperature effective
potential. For a given model, it is given by
\begin{equation}
\mathcal{F}(\phi,T)=V(\phi) +V_T(\phi),  \label{ftot}
\end{equation}%
where $V(\phi)$ is the zero-temperature effective potential and $V_T(\phi)$
the finite-temperature correction. To one-loop order, the latter is given by
\cite{quiros}
\begin{equation}
V_T(\phi) =\sum_{i}(\pm {g_{i}})T
\int\frac{d^3p}{(2\pi)^3}\log \left( 1\mp e^{-E_i/T} \right),
\label{f1loop}
\end{equation}
where the sum runs over particle species, $g_{i}$ is the number of degrees
of freedom of species $i$, the upper sign stands for bosons, the lower sign
stands for fermions, and $E_i=\sqrt{p^2+m_i^2(\phi)}$, where $m_i$ are the
Higgs-dependent masses.

We may have a phase transition if the  free-energy density has two minima $\phi
_{\pm }(T)$, corresponding to the two phases of the system. At high
temperatures the absolute minimum is  $\phi _{+}$, while at low temperatures
the absolute minimum is $\phi _{-}$. Hence, the system is initially in a state
characterized by $\phi (x)\equiv \phi _{+}$, which we shall refer to as ``the
$+$ phase''. Similarly, at late times the universe is in ``the $-$ phase'',
characterized by $\phi (x)\equiv \phi _{-}$. In the case of a first-order phase
transition, there is a temperature range in which these minima coexist in the
free energy, separated by a barrier. The critical temperature $T_{c}$ is given
by the condition $ \mathcal{F}(\phi _{+},T_{c})=\mathcal{F}(\phi _{-},T_{c})$.
Below the critical temperature, bubbles of the $-$ phase appear, inside which
we have $\phi =\phi _{-}$.

The growth of a bubble can be studied by considering the equations for the
variables $\phi,u^\mu,T$. The dynamics of the fluid variables can be
obtained from the conservation of the energy-momentum tensor. For the
Higgs-fluid system we have (see, e.g. \cite{ikkl94})
\begin{equation}
T_{\mu \nu }=\partial _{\mu }\phi \partial _{\nu }\phi -g_{\mu \nu }\left[
\frac{1}{2}\partial _{\alpha }\phi \partial ^{\alpha }\phi -\mathcal{F}(\phi
,T)\right] -u_{\mu }u_{\nu }T\frac{\partial \mathcal{F}}{\partial T}(\phi
,T),
\end{equation}
with $g^{\mu \nu }=\mathrm{diag}(1,-1,-1,-1)$. Conservation of $T^{\mu \nu
}$ gives
\begin{equation}
\partial _{\mu }\left[ T\frac{\partial \mathcal{F}}{\partial T}u^{\mu
}u^{\nu }-\mathcal{F}g^{\mu \nu }\right] =\square \phi \,\partial ^{\nu
}\phi .  \label{constmn}
\end{equation}
These equations govern the fluid dynamics and also contain the interaction
of the fluid with the scalar field $\phi$. The evolution of $\phi$ is
governed by a finite-temperature equation of motion of the form \cite{mp95}
\begin{equation}
\square \phi +\frac{\partial \mathcal{F}}{\partial \phi }+
\sum_i g_i\frac{dm^2_i}{d\phi}\int\frac{d^3p}{(2\pi)^32E_i}
\delta f_i=0.
\label{fieldeq}
\end{equation}
Here, the derivative of the finite-temperature effective potential takes into
account quantum and thermal corrections to the tree-level field equation, where
the thermal corrections are calculated from the equilibrium distribution
functions $f_i^{\mathrm{eq}}(p)=1/(e^{E_i/T}\mp 1)$. On the other hand, $\delta
f_i$ are the deviations from the equilibrium distributions. The last term
constitutes a damping due to the presence of the plasma. Computing $\delta f_i$
generally involves solving a system of Boltzmann equations which take into
account all the particles interactions.

In the bubble configuration, the bubble wall separates the two phases, i.e.,
the regions with $\phi=\phi_+$ and $\phi=\phi_-$. Thus, by definition, the
field varies only inside the bubble wall. As a consequence, away from the wall,
Eqs.~(\ref{constmn}) give equations for the fluid alone,
\begin{equation}
\partial _{\mu
}T_{\mathrm{fl}}^{\mu \nu }=0,\quad \mbox{with}\quad T_{\mathrm{fl}}^{\mu \nu }=u^{\mu }
u^{\nu }w-g^{\mu \nu }p,  \label{tmnp}
\end{equation}
where $p$ is the pressure and $w$ is the enthalpy density. In each phase, these
quantities are given by the free-energy density
$\mathcal{F}_\pm(T)=\mathcal{F}(\phi_\pm,T)$ through the well-known
thermodynamic relations $p=-\mathcal{F}, w=Tdp/dT=e+p$, where $e$ is the energy
density. The energy involved in the wall and fluid motions and in the reheating
comes from the difference of energy density between the two phases. This energy
is released at the phase transition fronts. The latent heat is defined as
$L=e_+(T_c)-e_-(T_c)$. For the treatment of hydrodynamics, the wall can be
assumed to be infinitely thin. Therefore, we shall simplify the system by
considering the fluid equations (\ref{tmnp}) together with an equation of
motion for the wall (rather than for the Higgs field).

An equation for the wall can be obtained from Eq.~(\ref{fieldeq}) by
multiplying by $\partial_\mu \phi$ and integrating in the direction
perpendicular to the wall (see Sec.~\ref{micro}). In this way, from the first
term in (\ref{fieldeq}) we obtain a term which is proportional to the wall
acceleration. If we ignore hydrodynamics (i.e., temperature gradients), the
second term gives the difference between the pressures on each side of the
wall, $p_--p_+$. This is a positive force acting on the wall. In contrast, the
deviations from equilibrium $\delta f_i$ in the last term turn out to oppose
the wall motion. It is well known that, for a small wall velocity $v_w$, the
last term in (\ref{fieldeq}) gives a term proportional to $-v_w$ in the wall
equation, i.e., a friction force. As a consequence of the friction, the wall
may reach a steady-state regime of constant velocity. However, it is known that
such a steady state does not always exist, either due to instabilities which
make the wall motion turbulent \cite{inst}, or just because the friction is not
high enough to prevent the wall to run away \cite{bm09}. In the latter case,
the wall quickly reaches velocities $v_{w}\simeq 1$, with increasingly high
values of the gamma factor.

Interestingly, the ultra-relativistic case turns out to be much simpler than
the non-relativistic one. This is because particles which cross the UR wall do
not have time to interact, and Boltzmann equations need not be considered. In
this case, it is simpler to compute the complete occupancies $f_i$ rather than
the deviations $\delta f_i$, i.e., to consider the second and third terms of
Eq.~(\ref{fieldeq}) simultaneously. Macroscopically, this amounts to
calculating the {total} force acting on the wall. The result (for particle
masses which vanish in the $+$ phase) is a net force given by \cite{bm09}
\begin{equation}
F_{\mathrm{net}}=V(\phi_+)-V(\phi_-)-\sum_i g_i c_i\frac{T_+^2m_i^2(\phi_-)}{24},
\label{Fnet}
\end{equation}
where $c_i=1$ ($1/2$) for bosons (fermions), and $T_+$ is the temperature of
the unperturbed fluid in front of the wall. Notice that this force does not
depend on the wall velocity. As a consequence, if the wall reaches the UR
regime with a positive $F_{\mathrm{net}}$, then it will run away.

In order to determine the actual value of $v_{w}$, the force acting on the wall
in the whole range $0<v_{w}<1$ is needed. The friction force seems to be
generally a growing function of $v_{w}$, although the usual NR approximations
break-down around the speed of sound \cite{knr14}. The fact that
$F_{\mathrm{net}}$ becomes independent of $v_{w}$ in the ultra-relativistic
limit implies that the friction force saturates as a function of $v_{w}\gamma
_{w}$. A phenomenological model for the friction force, which interpolates
between the NR and UR regimes, was introduced in Ref. \cite{ariel13}. It
consists in replacing the last term in the field equation (\ref{fieldeq}) with
a simpler damping term,
\begin{equation}
\mathcal{K}=\frac{f(\phi )\,u^{\mu }\partial _{\mu }\phi }{\sqrt{1+[g(\phi
)\,u^{\mu }\partial _{\mu }\phi ]^{2}}}, \label{Kfeno}
\end{equation}
where $f$ and $g$ are  scalar functions which can be chosen suitably to give
the correct $\phi$ dependence of the friction. Considering Eq.~(\ref{Kfeno}) in
the wall frame, it was shown in \cite{ariel13} that this term gives a friction
force which has the correct velocity dependence in the NR and UR limits. In
Sec.~\ref{micro} we shall repeat the derivation in the plasma frame.

\section{Hydrodynamics}  \label{hidro}

Let us consider a wall moving with velocity $v_w$.  We will assume that the
wall is infinitely thin, and that the symmetry of the problem is such that the
velocity of the fluid is perpendicular to the wall  (e.g., spherical or planar
symmetry). Thus, the fluid is characterized by two variables, namely,  the
temperature $T$ and a single component of the velocity, $v$. We are interested
in supersonic walls, i.e., with $v_w>c_+$, where $c_+=\sqrt{dp_+/de_+}$ is the
speed of sound in the plasma in the $+$ phase. Concretely, we shall only
consider wall velocities which are so high that the fluid in front of the wall
is unperturbed. Therefore, in the $+$ phase the fluid velocity $v_+$ vanishes
and the temperature $T_{+}$ is set by the nucleation temperature. We will also
assume that the fluid behind the wall is in local equilibrium, so that the
variables $T$ and $v$ are well defined everywhere. Inside the bubble, the fluid
variables are given by Eqs.~(\ref{tmnp}). Their values $v_-,T_-$ next to the
wall can be obtained by integrating the equations $\partial _{\mu }T^{\mu \nu
}=0$ across the interface. Besides, we have the boundary condition that the
fluid velocity vanishes at the bubble center.

\subsection{Detonations}

In the steady-state case, it is usual to consider the reference frame of the
wall, so that time derivatives vanish. We shall consider instead the rest frame
of the unperturbed fluid in front of the wall (the plasma frame), where it is
easier to take the limit $v_{w}\rightarrow 1$. We thus have $v_+=0$. We can
obtain $v_{-}$, $T_{-}$ as functions of $v_{w},T_{+}$ from the continuity
equations for energy and momentum.

Consider a piece of wall of surface area $A$ which is small enough that it can
be regarded as planar. Locally, we place the coordinate system so that the wall
moves in the positive $z$-direction with velocity $v_{w}$. Thus, we only need
to consider time and $z$ components of $T^{\mu \nu }$. In a small time $\Delta
t$ the wall moves a distance $\Delta z=v_{w}\Delta t$. During this time, we
have an incoming energy flux from the left of the wall, given by $T_{-}^{0z}$,
and an outgoing flux to the right, given by $T_{+}^{0z}$. Therefore, a net
energy $(T_{-}^{0z}-T_{+}^{0z})A\Delta t$ will accumulate in the interface
unless it is transferred to the plasma. The change of energy in the plasma in
the volume $A\Delta z$ is given by $(T_{-}^{00}-T_{+}^{00})Av_w\Delta t$. For a
steady-state wall the energy balance gives
\begin{equation}
T_{-}^{0z}-T_{+}^{0z}=(T_{-}^{00}-T_{+}^{00})v_{w}. \label{enbal}
\end{equation}
Similarly, considering the momentum density $T^{z0}$ and momentum flux $
T^{zz}$, we obtain
\begin{equation}
T_{-}^{zz}-T_{+}^{zz}=(T_{-}^{z0}-T_{+}^{z0})v_{w}. \label{mombal}
\end{equation}

On each side of the wall, we have $T^{\mu \nu }=T_{\mathrm{fl} }^{\mu \nu }$
given by Eq.~(\ref{tmnp}). In our reference frame we have $u^{\mu
}=(\gamma,0,0,\gamma v)$, with
$\gamma=1/\sqrt{1-v^2}$. Since $v_+=0$ and $v_->0$, we have 
\begin{equation}
T_{+}^{00}=e_{+},\quad T_{+}^{0z}=T_{+}^{z0}=0,\quad T_{+}^{zz}=p_{+},
\label{tmas}
\end{equation}
and 
\begin{equation}
T_{-}^{00}=w_{-}\gamma _{-}^{2}-p_{-},\quad
T_{-}^{0z}=T_{-}^{z0}=w_{-}\gamma _{-}^{2}v_{-},\quad
T_{-}^{zz}=w_{-}\gamma _{-}^{2}v_{-}^{2}+p_{-}.  \label{tmenos}
\end{equation}
Inserting Eqs.~(\ref{tmas}-\ref{tmenos}) in (\ref{enbal}-\ref{mombal}), we
obtain the system of equations
\begin{eqnarray}
w_{-}(v_{w}-v_{-}) &=&(p_{-}+e_{+})v_{w}(1-v_{-}^{2}), \label{junc1} \\
w_{-}(v_{w}-v_{-})v_{-} &=&(p_{-}-p_{+})(1-v_{-}^{2}), \label{junc2}
\end{eqnarray}
from which we readily obtain
\begin{equation}
v_{-}v_{w}=\frac{p_{-}-p_{+}}{e_{+}+p_{-}},\quad \frac{v_{-}}{v_{w}}=\frac{
e_{-}-e_{+}}{e_{-}+p_{+}}.  \label{velplasma}
\end{equation}
These expressions are similar, but different, to the usual expressions for
$v_{+}$ and $v_{-}$ in the wall frame.

Since the variables $w,p,e,T$ are related by the equation of state (EOS),
Eqs.~(\ref{velplasma}) can be solved for, say, $w_-$ and $v_-$ as functions of
$w_+$ and $v_w$. It is not difficult to see that the derivatives $\partial
w_-/\partial v_w|_{T_+}$ and $\partial v_-/\partial v_w|_{T_+}$ diverge for
$v_w$ such that\footnote{This can be seen by differentiating
Eqs.~(\ref{junc1}-\ref{junc2}) for fixed $T_+$ and using the relation
$dp_-=dw_-/(1+c_-^2)$ (see also \cite{mm14deto}).}
\begin{equation}
\frac{v_w-v_-}{1-v_wv_-}= c_-, \label{Jouguet}
\end{equation}
where $c_-=\sqrt{dp_-/de_-}$ is the speed of sound in the $-$ phase. 
The left-hand side of Eq.~(\ref{Jouguet}) gives the value of the fluid velocity in the
reference frame of the wall. Therefore, the mentioned divergence occurs when
the outgoing flow velocity in the wall frame reaches the speed of sound. This
indicates that the hydrodynamics becomes too strong at this point. For a
detonation, the incoming flow velocity is given by $v_w$, which is supersonic.
Detonations are divided into weak detonations, for which the outgoing flow is
supersonic too, strong detonations, for which the outgoing flow is subsonic,
and Jouguet detonations, which are characterized by the condition
(\ref{Jouguet}). This means that in the plasma frame, weak detonations
correspond to smaller values of $v_-$ (i.e., to solutions which do not perturb
the fluid strongly) while strong detonations correspond to higher values of
$v_-$.

\begin{figure}[bt]
\centering
\epsfxsize=7.5cm \leavevmode \epsfbox{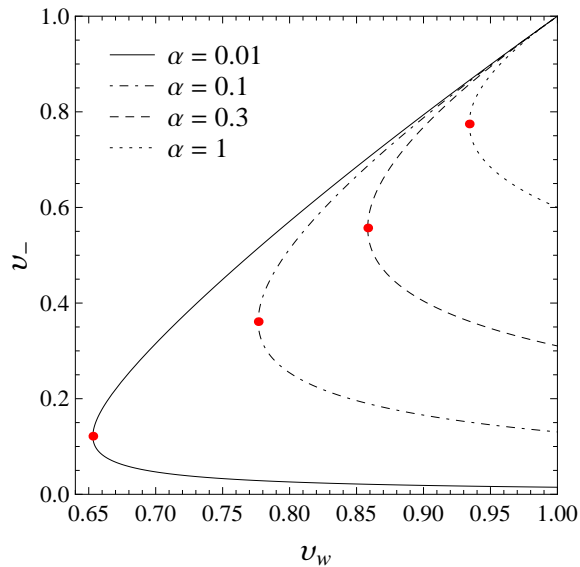}
\caption{
The fluid velocity behind the wall as a function of the wall velocity for
the bag EOS (see below), for several values of $\alpha=L/(3w_+)$.}
\label{figvmedet}
\end{figure}

The behavior of $v_-$ as a function of $v_w$ is shown in Fig.~\ref{figvmedet}
for a simple equation of state (the bag EOS) considered below. The upper branch
corresponds to strong detonations, the lower branch corresponds to weak
detonations, and the red dots indicate the Jouguet point.
The abovementioned divergence of $\partial v_-/\partial v_w|_{T_+}$ can be
observed in the figure. It causes $v_w$ to be a minimum at the Jouguet point.
As a consequence, for detonations the wall velocity is in the range $v_{J}\leq
v_{w}<1$, where $v_{J}(T_+)$ is velocity of the Jouguet detonation. It is well
known that strong detonations are not compatible with the solutions for the
fluid profile behind the wall. Therefore, the upper curves in
Fig.~\ref{figvmedet} do not correspond to physical solutions. Weak detonations
become weaker (i.e., $v_-$ decreases) for higher wall velocities. In the limit
$v_w\to 1$, Eqs.~(\ref{velplasma}) become
\begin{equation}
e_{-}-p_{-}=e_{+}-p_{+},\quad v_{-}=\frac{p_{-}-p_{+}}{e_{+}+p_{-}}=\frac{
e_{-}-e_{+}}{e_{+}+p_{-}}.  \label{hydreldet}
\end{equation}
The first of these equations gives the temperature $T_{-}$ as a function of $
T_{+}$ for an ultra-relativistic stationary solution. The second one gives the
fluid velocity behind the interface.

\subsection{Runaway walls}

If the wall is accelerated, we have to take into account the fact that a part
of the energy accumulates in the wall \cite{bm09}. In the time $\Delta t$, an
amount of energy $A\Delta \sigma $ is accumulated in a surface area $A$ of the
interface, where $\sigma $ is the surface energy density. Hence, the energy
balance now gives
\begin{equation}
T_{-}^{0z}-T_{+}^{0z}=(T_{-}^{00}-T_{+}^{00})v_{w}+\frac{d\sigma }{dt}.
\label{juncr1}
\end{equation}
Similarly, since the momentum of a piece of wall is given by $A\sigma v_{w} $,
we have
\begin{equation}
T_{-}^{zz}-T_{+}^{zz}=(T_{-}^{z0}-T_{+}^{z0})v_{w}+\frac{d(v_{w}\sigma )}{dt}.
\label{juncr2}
\end{equation}

After a certain  (generally short) period of time, the accelerated wall will
either reach a  terminal velocity or accelerate to ultra-relativistic
velocities. The ultra-relativistic accelerated regime is similar to the 
steady-state case in the sense that the wall velocity is essentially a constant,
$v_{w}\simeq 1$ (although $\gamma_w$ and $\sigma$ vary). In this limit we have
\begin{equation}
\frac{d\sigma }{dt}=\frac{d(v_{w}\sigma )}{dt}=F_{\mathrm{net}},
\label{dsigf}
\end{equation}
where $F_{\mathrm{net}}$ is the net force per unit area acting on the wall.
Inserting Eqs.~(\ref{tmas}-\ref{tmenos}) in (\ref{juncr1}-\ref{juncr2})  we
obtain
\begin{equation}
e_{-}-p_{-}=e_{+}-p_{+}-2F_{\mathrm{net}},\quad v_{-}=\frac{w_{-}-w_{+}}{
w_{-}+w_{+}}.  \label{eqsrun}
\end{equation}
For $F_{\mathrm{net}}=0$, Eqs.~(\ref{eqsrun}) match the ultra-relativistic
detonation case, Eq.~(\ref{hydreldet}). From Eqs.~(\ref{eqsrun}) we may obtain
the temperature $T_{-}$ and the velocity $v_{-}$ as functions of
$F_{\mathrm{net}}$ and $T_{+}$. We see that, for a constant net force, $v_-$
and $T_-$ are constant, like in the stationary case.

\subsection{Fluid profiles}

The profiles of $v$ and $T$ behind the wall are a solution of Eqs.~(\ref{tmnp})
with boundary conditions $v=v_{-}$, $T=T_{-}$ at the  wall. For a system with
spherical, cylindrical or planar symmetry, the problem is 1+1 dimensional,
since the fluid profile depends only on time and on the distance $r$ from the
center, axis or plane of symmetry \cite{kurki85}. Besides, since there is no
distance scale in the fluid equations, it is customary to assume the similarity
condition, namely, that the solutions depend only on the variable $\xi =r/t$.
With this assumption, one obtains the equation for the wall velocity
\cite{lm11}
\begin{equation}
\gamma ^{2}(1-v\xi )\left[ \frac{1}{c_-^{2}}\left( \frac{\xi -v}{1-\xi v}
\right) ^{2}-1\right] v^{\prime }=j\frac{v}{\xi },  \label{fluideqv}
\end{equation}
where a prime indicates a derivative with respect to $\xi$, and $j=2$, $1$, or
$0$ for  spherical, cylindrical, or planar walls, respectively. The enthalpy
profile is given by the equation
\begin{equation}
\frac{w^{\prime }}{w}=\left( \frac{1}{c_-^{2}}+1\right) \frac{\xi -v}{
1-\xi v}\gamma ^{2}v^{\prime }.  \label{fluideqw}
\end{equation}

It is important to note that the similarity condition is compatible with a wall
which is placed at a fixed value of $\xi$, namely, $\xi _{w}=v_{w}$. For an
accelerated wall, this condition will not be compatible, in general, with the
boundary conditions at the interface. Nevertheless, in the ultra-relativistic
limit, the wall position corresponds essentially to the constant value $\xi
_{w}=1$. Indeed, as we have seen, the values of $ T_{-}$ and $v_{-}$ are
constant in this limit for a constant $F_{\mathrm{net}}$. Therefore, the fluid
profiles for the runaway solution can be obtained from
Eqs.~(\ref{fluideqv}-\ref{fluideqw}), like in the detonation case.

\subsection{The bag EOS}

In order to  solve the hydrodynamic equations we need to consider a particular
equation of state. We shall consider the bag EOS, in which the two phases
consist  of radiation and vacuum energy. This approximation has been widely used
for simplicity, and also in order to obtain model-independent results which
depend on a few physical quantities. Setting the vacuum energy in the
low-temperature phase to zero, the model depends on three physical parameters,
which we may choose to be the critical temperature $T_{c}$, the latent heat
$L$, and the radiation constant of the high-temperature phase,  $a$. Thus, we
write
\begin{equation}
p_{+}(T)=\frac{1}{3}aT^{4}-\frac{L}{4},\quad p_{-}(T)=\frac{1}{3}\left( a-\frac{
3L}{4T_{c}^{4}}\right) T^{4}.
\end{equation}
The energy density of the high-temperature phase is of the form $
e_{+}(T)=aT^{4}+\varepsilon $, where the false-vacuum energy density is given
by $\varepsilon =L/4$. In the low-temperature phase, the energy density is of
the form $e_{-}(T)=a_{-}T^{4}$, with a radiation constant given by $a_{-}=a(1-
3\alpha_{c})$, where
\begin{equation}
\alpha_{c}=\varepsilon/(aT_{c}^{4}).
\end{equation}
We define the usual bag variable
\begin{equation}
\alpha\equiv\varepsilon/(aT_+^4)=L/(3{w_{+}}). \label{Lbar}
\end{equation}
The enthalpy density is given by  $w_{\pm}=(4/3)a_\pm T_{\pm}^{4}$ (with
$a_+\equiv a$). For the bag EOS the speed of sound is the same in both phases,
$c_{\pm}=1/\sqrt{3}$.

From Eqs.~(\ref{velplasma}) we obtain the fluid variables behind a weak
detonation wall,
\begin{eqnarray}
v_{-}&=&\frac{3\alpha-1+3(1+\alpha)v_{w}^{2}-\sqrt{\left[ 3\alpha-1+3(1+\alpha
)v_{w}^{2}\right] ^{2}-12\alpha(2+3\alpha)v_{w}^{2}}}{2(2+3\alpha)v_{w}},
\label{vme} \\
\frac{w_{-}}{w_{+}}&=&\frac{\gamma _{w}^{2}}{3}\left[ 1-3\alpha+3(1+\alpha
)v_{w}^{2}-2\sqrt{\left( 1-3\alpha+3(1+\alpha)v_{w}^{2}\right)
^{2}-12v_{w}^{2}}\right]   \label{wmewma}
\end{eqnarray}
(there is also a solution with a $+$ sign in front of the square roots,
corresponding to strong detonations). Notice that the fluid velocity and the
enthalpy ratio depend only on the variable $\alpha$ and the wall velocity. On
the other hand, the temperature is given by
\begin{equation}
\frac{T_{-}^{4}}{T_{+}^{4}}=\frac{1}{1-3\alpha_{c}}\,\frac{w_{-}}{w_{+}}.
\label{tmetmabag}
\end{equation}
The Jouguet velocity is obtained by considering the condition (\ref{Jouguet})
together with Eqs.~(\ref{velplasma}). For the bag EOS we obtain
\begin{equation}
v_{J}=\frac{\sqrt{2\alpha+3\alpha^{2}}+1}{\sqrt{3}(1+\alpha)}.
\end{equation}
The ultra-relativistic limit can be obtained either by taking the limit
$v_{w}\rightarrow 1$ in Eqs.~(\ref {vme}-\ref{wmewma}), or directly from
(\ref{hydreldet}). We have
\begin{equation}
\frac{w_{-}}{w_{+}}=1+3\alpha,\quad v_{-}=\frac{3\alpha}{2+3\alpha}\quad
\mbox{(UR detonation).}  \label{tmevmeurdeto}
\end{equation}

\begin{figure}[bth]
\centering
\epsfxsize=15cm \leavevmode \epsfbox{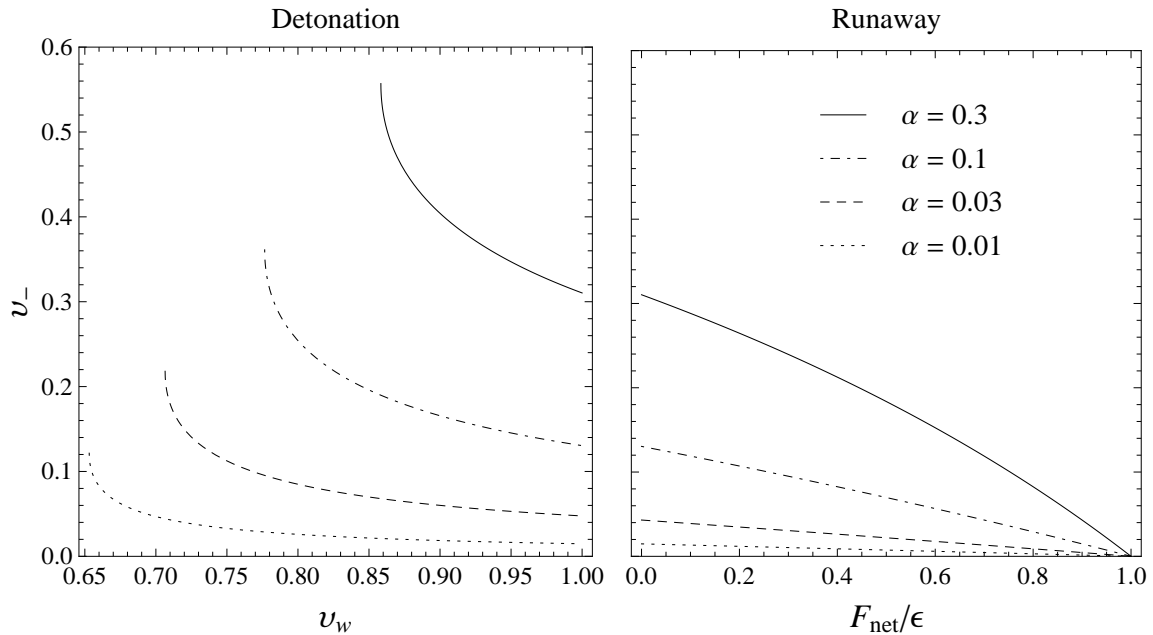}
\caption{
The fluid velocity behind the wall as a function of the wall velocity for
the case of a detonation (left panel) and as a function of the net force for
the case of a runaway wall (right panel).}
\label{figvmetme}
\end{figure}

For a runaway wall we obtain, from Eqs.~(\ref{eqsrun}),
\begin{equation}
\frac{w_{-}}{w_{+}}=1+3(\alpha-\bar{F}),\quad v_{-}=\frac{3(\alpha-\bar{F})}{
2+3(\alpha-\bar{F})}\quad \mbox{(runaway)},  \label{tmevmerun}
\end{equation}
where
\begin{equation}
\bar{F}\equiv \frac{F_{\mathrm{net}}}{aT_+^4}=\frac{4}{3}\frac{F_{\mathrm{net}}}{w_+}.
\label{Fbar}
\end{equation}
Notice that Eqs.~(\ref{tmevmerun}) match the detonation case for
$F_{\mathrm{net}}=0$. On the other hand, as $F_{\mathrm{net}}$ increases,
$T_{-}$ and $v_{-}$ decrease. Hence, the hydrodynamics becomes weaker for
larger acceleration. This behavior is similar to the detonation case, in which
the higher the wall velocity, the weaker the hydrodynamics (see
Fig.~\ref{figvmetme}). This is related to the fact that the hydrodynamics
obstructs the wall motion \cite{ms09,kn11}. Moreover, we see that for $\bar
F=\alpha$ we have $w_{-}=w_{+}$ and $v_{-}=0$, i.e., the fluid remains
unperturbed after the passage of the wall.

The condition $\bar F=\alpha$ sets a maximum value for the net force, which is
given by the false vacuum energy density, $F_{\mathrm{\max }}=\varepsilon$. To
understand this physically, notice that the force which drives the wall motion
is essentially given by the pressure difference between the two phases. This
force vanishes at the critical temperature and reaches its maximum at zero
temperature. At $T=0$ the pressure difference is just given by the
zero-temperature effective potential, and coincides with the false-vacuum
energy density. In the bag model, this is given by the parameter $\varepsilon $
(at finite temperature there is also a friction force due to the plasma, but at
zero temperature the friction force vanishes). Therefore, $F_{\mathrm{net}}$
can reach the maximum value $\varepsilon$ if the phase transition occurs at
$T_{+}=0$. However, such an extreme supercooling is not likely in concrete
physical models.

\begin{figure}[bth]
\centering
\epsfysize=7cm \leavevmode \epsfbox{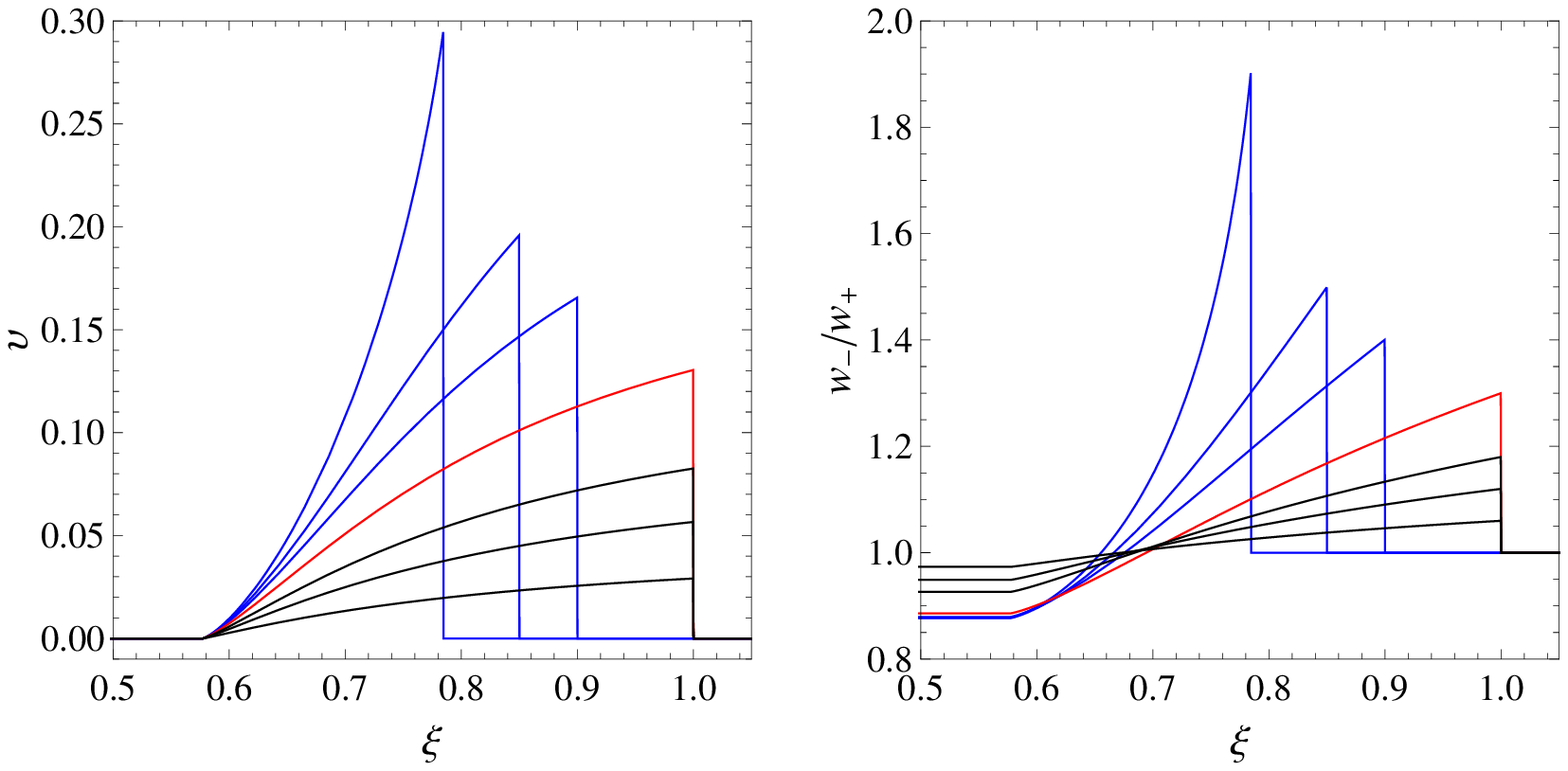}
\caption{Fluid velocity and enthalpy profiles for spherical bubbles, for
$\alpha=0.1$ and different wall velocities.
The highest curve corresponds to a Jouguet detonation with
wall velocity $v_{J}(\alpha)\simeq 0.78$.
The other blue curves correspond to weak detonations with velocities $
v_{w}=0.85$ and $v_{w}=0.92$. The red curve corresponds to the limit of a
detonation with $v_{w}\rightarrow 1$ or a runaway wall with $F_{\mathrm{net}
}\rightarrow 0$. The black curves correspond to runaway walls with $F_{
\mathrm{net}}/F_{\max }=0.4$, $0.6$, and $0.8$.}
\label{figprof}
\end{figure}

For the bag EOS it is relatively simple to obtain the fluid profiles, since
$c_-$ is a constant. However, except for the planar case, the fluid equations
(\ref{fluideqv}-\ref{fluideqw}) must be solved numerically. Behind the wall,
the solutions which fulfil the boundary condition of a vanishing fluid velocity
at $\xi=0$ are  rarefaction waves, in which $v(\xi)$ actually vanishes for
$0<\xi <c_{-}$ and grows for $\xi>c_-$ up to the boundary value $v_{-}$ at
$\xi=\xi_w$ (see e.g. \cite{ekns10,lm11}). The temperature and pressure also
decrease away from the wall. These can be computed from the enthalpy profile,
which is readily obtained by integrating Eq.~(\ref{fluideqw}),
\begin{equation}
\frac{w}{w_-}=\exp\left[\left(\frac{1}{c_-^2}+1\right)\int_{v_-}^{v}
\frac{\xi-v}{1-\xi v}\gamma^2\,dv\right].  \label{entprof}
\end{equation}
In Fig.~\ref{figprof} we show some profiles for the case of
spherically-symmetric bubbles.

\subsection{Efficiency factors}

For the bag EOS, the efficiency factor is defined as the fraction of the
released vacuum energy $\varepsilon $ which goes into bulk motions of the
fluid\footnote{The total released energy density at finite temperature, $\Delta
e(T_{+})$, is actually higher than $\varepsilon $. At $T=T_{c}$ we have $\Delta
e=L=4\varepsilon $, while  $\Delta e=\varepsilon $ occurs only at $T=0$. See
\cite{lm15} for an alternate definition of an efficiency factor.},
\begin{equation}
\kappa_{\mathrm{fl}} \equiv\frac{E_{\mathrm{kin}}}{\varepsilon V_{b}},
\end{equation}
where $V_b$ is the volume of the bubble. We have, for the different wall
symmetries \cite{lm11},
\begin{equation}
\kappa_{\mathrm{fl}}=\frac{j+1}{\varepsilon
v_{w}^{j+1}}\int_{0}^{\infty }\!d\xi \,\xi ^{j}\,w\,\gamma ^{2}v^{2}.
\end{equation}
As can be seen from Eq.~(\ref{entprof}), for detonations or runaway walls the
profile of $w/w_-$ does not depend on $w_-$ but only on the profile of $v$,
which only depends on the boundary values $v_w,v_{-}$. As a consequence, we can
write
\begin{equation}
\kappa _{\mathrm{fl}}=
\frac{j+1}{v_{w}^{j+1}}\,\frac{4}{3\alpha}\frac{w_{-}}{w_{+}}
\,I(v_{w},v_{-}),  \label{kappa}
\end{equation}
with
\begin{equation}
I(v_{w},v_{-})=
\int_{c_{-}}^{v_{w}}\!d\xi \,\xi ^{j}\,\frac{w}{w_{-}}\,\gamma ^{2}v^{2}.
\label{integ}
\end{equation}
For detonations, $v_-$ and $w_-/w_+$ are functions of $\alpha$ and $v_w$, and
therefore the efficiency factor depends only on these quantities,
$\kappa_{\mathrm{fl}} =\kappa_{\mathrm{fl}} ^{\mathrm{det}}(\alpha,v_{w})$. On
the other hand, for runaway walls we have $v_w=1$, while $w_-/w_+$ and $v_{-}$
depend on $\alpha$ and $\bar F$. From Eqs.~(\ref{kappa}) and (\ref{tmevmerun}),
in this case $\kappa_{\mathrm{fl}}$ is of the form
\begin{equation}
\kappa _{\mathrm{fl}}^{\mathrm{run}}(\alpha,\bar{F})=\frac{j+1}{3}\,\frac{4(1+3\alpha-3\bar{F})}{\alpha
}\,I_{1}(v_{-}),  \label{kapparun}
\end{equation}
where $I_{1}(v_{-})=I(1,v_{-})$.

\begin{figure}[bth]
\centering
\epsfxsize=15cm \leavevmode \epsfbox{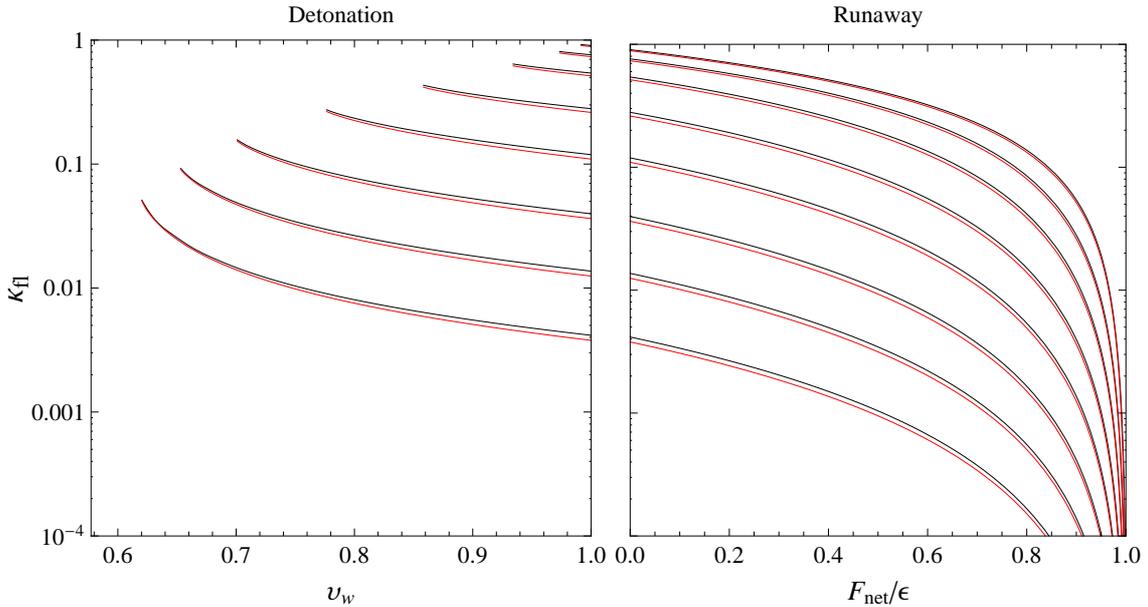}
\caption{The efficiency factor for spherical (black) and planar (red) walls,
for several values of $\alpha$. From bottom to top,
$\protect\alpha =0.003,0.01,0.03,0.1,0.3,1,3,10$.}
\label{figplesf}
\end{figure}

For the planar case the integral (\ref{integ}) can be done analytically
\cite{lm11}, while for spherical or cylindrical walls it must be calculated
numerically. Notice that $w_-$ and $v_-$ are the same for all these cases, but
the rarefaction profiles differ. In Fig.~\ref{figplesf}  we show the value of
the efficiency factor for spherical and planar walls. The cylindrical case lies
between the other two. For steady-state walls we plotted $\kappa
_{\mathrm{fl}}$ as a function of the wall velocity (left panel), while for
runaway walls we plotted it as a function of the net force (right panel).
For the range of values shown in the figure, the difference between the two
wall symmetries is always less than a 10\%, while this relative difference is
exceeded only for $F_{\mathrm{net}}$ very close to $F_{\max }=\varepsilon$,
where $\kappa _{\mathrm{fl}}\rightarrow 0$. As already discussed, it is not
likely that such values of $F_{\mathrm{net}}$ will be reached in a physical
model. Such a small difference is interesting, since for planar walls we obtain
analytic results (see the appendix). In the appendix we also give fits for
spherical detonations and runaway walls, where the relative error is smaller
than a 3\% in the whole detonation range and in most of the runaway range.

\begin{figure}[bth]
\centering
\epsfxsize=15cm \leavevmode \epsfbox{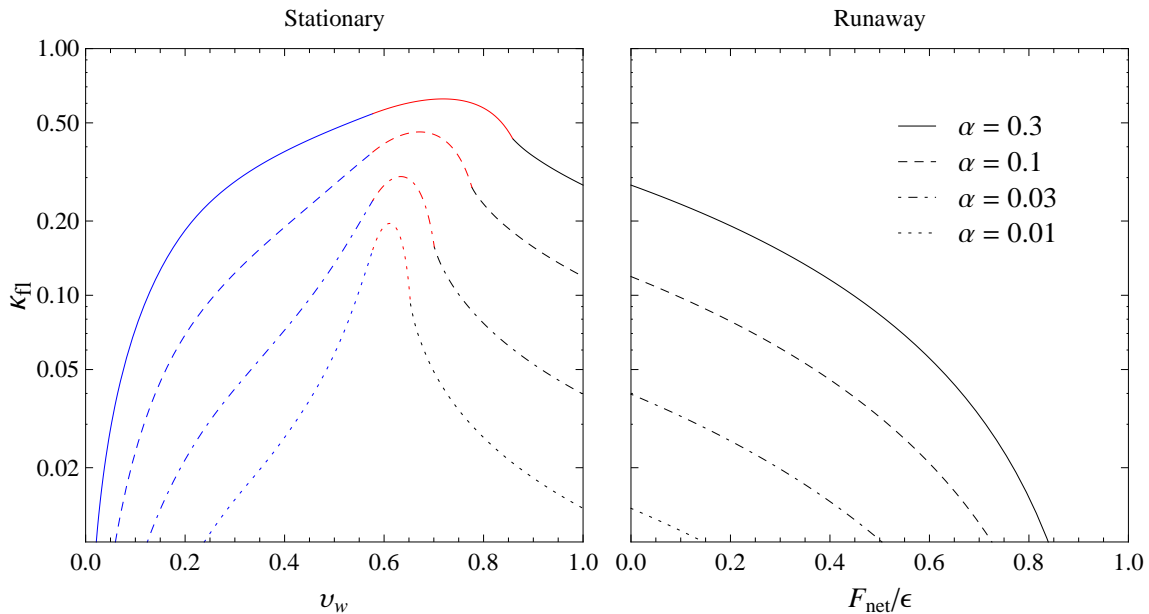}
\caption{The efficiency factor $\protect
\kappa _{\mathrm{fl}}$ for spherical bubbles. Blue curves correspond to weak deflagrations,
red curves to Jouguet deflagrations, and black curves to weak detonations
or runaway walls.}
\label{figkesf}
\end{figure}

For comparison, we show in Fig.~\ref{figkesf} the value of
$\kappa_{\mathrm{fl}}$  in the different regimes,  for some values of the bag
parameter $\alpha $ (for the calculation in the cases of weak and Jouguet
deflagrations, see \cite{lm11}). The different types of stationary solutions
are divided by the points $v_w=c_-$ and $v_w=v_J(\alpha)$.
As already discussed, the hydrodynamics of weak detonations becomes weaker as
the wall velocity  increases, and it becomes even weaker for runaway walls.
This is reflected in the efficiency factor, which is monotonically decreasing
in these regimes. On the other hand, for small wall velocities (i.e., for weak
deflagrations) the efficiency factor increases with the wall velocity, and is
maximal for supersonic (Jouguet) deflagrations.

In the runaway case, a portion of the released vacuum energy goes into kinetic
energy of the wall. This will increase the efficiency for gravitational wave
generation through direct bubble collisions. This fraction is given by
\begin{equation}
\kappa_{\mathrm{wall}}\equiv\frac{\Delta E_{\mathrm{wall}}}{\varepsilon \Delta V_{b}}.
\end{equation}
Considering a small piece of the thin interface, the energy which goes to the
corresponding volume $\Delta V_b=Av_w\Delta t$ is given by $\Delta
E_{\mathrm{wall}}=\Delta \sigma A$. Therefore, we have, from Eq.~(\ref{dsigf})
and the definitions (\ref{Lbar},\ref{Fbar}),
\begin{equation}
\kappa_{\mathrm{wall}}=\frac{d\sigma /dt}{
\varepsilon v_{w}}
=\frac{F_{\mathrm{net}}}{\varepsilon}=\frac{\bar F}{\alpha}.  \label{enwall}
\end{equation}

\section{The wall equation of motion \label{micro}}

We shall now consider the equation of motion for the wall. At a given point of
the thin interface we may place the $z$ axis perpendicular to the wall. Then
the field equation (\ref{fieldeq}) with the damping (\ref{Kfeno}) becomes
\begin{equation}
(\partial _{0}^{2}-\partial _{z}^{2})\phi +\partial \mathcal{F}/\partial
\phi +\mathcal{K}=0 . \label{fieldeq0z}
\end{equation}
If we describe the wall at rest by a certain field profile $\phi _{0}(z)$ which
varies between the values $\phi _{-}$ and $\phi _{+}$, then, in the plasma
frame, we have
\begin{equation}
\phi(z,t) =\phi _{0}(\gamma _{w}\left(z-z_{w}\right)), \label{fieldprof}
\end{equation}
where $z_w$ depends on time and we have $v_{w}=\dot{z}_{w}$, $ \gamma
_{w}=1/\sqrt{1-v_{w}^{2}}$. The wall corresponds to the range of $z$ where
$\phi $ varies, and we define the wall position $z_{w}(t)$ by the condition
$\int (\partial_z\phi )^{2}(z-z_{w})\,dz\,=0$. Multiplying
Eq.~(\ref{fieldeq0z}) by $\partial_z\phi $ and integrating across the wall, we
obtain the equation
\begin{equation}
\sigma _{0}\gamma _{w}^{3}\dot{v}_{w}=\int_{-}^{+}\frac{\partial \mathcal{F}
}{\partial \phi }\,\frac{\partial \phi }{\partial z}\,dz+\int_{-}^{+}\mathcal{K}
\,\frac{\partial \phi }{\partial z}\,dz,  \label{eqwall}
\end{equation}
where $\int_{-}^{+}dz$ means integration between  points on each side of the
wall (where $\partial_z\phi $ vanishes) and
\begin{equation}
\sigma _{0}=\int_{-}^{+}\left[ \phi _{0}^{\prime }(z)\right] ^{2}dz.
\end{equation}
This integral gives the surface energy density of the wall at rest. We see that
it appears in the left-hand side of Eq.~(\ref{eqwall}) multiplying the proper
acceleration $\gamma _{w}^{3}\dot{v}_{w}$. Hence, the terms in the right-hand
side are forces (per unit area) acting on the wall. Since the force which
drives the wall motion is finite, the factor $\gamma _{w}^{3}$ implies that, in
the plasma frame, $\dot v_w$ decreases as the wall approaches the speed of
light.

The force terms in  Eq.~(\ref{eqwall}) depend not only on the Higgs profile but
also on the fluid profiles. The first of them is very sensitive to
hydrodynamics. For a constant temperature, this term gives the pressure
difference $p_{-}-p_{+}$. Since it is always positive, we shall refer to it as
the driving force $F_{\mathrm{dr}}$. The term containing $\mathcal{K}$
represents the microscopic departures from equilibrium caused by the moving
wall. It is always negative and velocity dependent. We shall refer to this term
as the friction force $F_{\mathrm{fr}}$. Thus, Eq.~(\ref{eqwall}) can be
written as
\begin{equation}
F_{\mathrm{net}}=F_{\mathrm{dr}}+F_{\mathrm{fr}} . \label{eqforces}
\end{equation}

\subsection{Driving force and hydrodynamic obstruction}

The driving force can be written in the form \cite{mm14deto}
\begin{equation}
F_{\mathrm{dr}}=p_{-}-p_{+}-\int_{-}^{+}\frac{\partial \mathcal{F}}{\partial
T^2}\,{dT^2},  \label{Fdr}
\end{equation}
which we shall approximate for definite calculations by
\begin{equation}
F_{\mathrm{dr} }\simeq p_{-}-p_{+}-\left\langle \frac{\partial \mathcal{F}}{\partial
T^{2}}\right\rangle (T_{+}^{2}-T_{-}^{2}), \label{Fdrapp}
\end{equation}
where we have approximated the value of ${\partial \mathcal{F}}/{\partial
T^{2}}$ inside the wall by
\begin{equation}
\left\langle \frac{\partial \mathcal{F}}{\partial
T^{2}}\right\rangle\equiv \frac{1}{2}\left(\frac{\partial \mathcal{F_+}}{\partial
T_+^{2}}+\frac{\partial \mathcal{F_-}}{\partial
T_-^{2}}\right).
\end{equation}
For the bag EOS, Eq. (\ref{Fdrapp}) takes the simple form
\begin{equation}
F_{\mathrm{dr}}=\frac{L}{4}\left( 1-\frac{T_{-}^{2}T_{+}^{2}}{T_{c}^{4}}
\right) = \varepsilon-\varepsilon \frac{T_{-}^{2}T_{+}^{2}}{T_{c}^{4}}.
\label{Fdrbag}
\end{equation}
In this approximation, the first term is the zero-temperature part of the
force. Indeed, as already mentioned, the false vacuum energy density is given
by the zero-temperature effective potential. Thus, for a physical model the bag
constant $\varepsilon$ would be given by $V(\phi_+)-V(\phi_-)$. The second term
in Eq.~(\ref{Fdrbag}) is the temperature-dependent part of the driving force,
and contains the effect of hydrodynamics. For weak detonations, this effect is
to increase $T_-$ with respect to the outside temperature $T_+$, and hence to
decrease the driving force. In this decomposition, the term $F_{\mathrm{hyd}}
=-(L/4) {T_{-}^{2}T_{+}^{2}}/{T_{c}^{4}}$ can be seen as a force which opposes
the wall motion and depends indirectly on the wall velocity (through the
dependence of the temperature on $v_{w}$). However, this hydrodynamic
obstruction does not behave as a fluid friction, since it \emph{decreases} with
the wall velocity. Indeed, the reheating behind the wall is highest for $v_w$
close to the Jouguet point and lowest for $v_{w}\rightarrow 1$. It is worth
remarking that the approximation (\ref{Fdrbag}) preserves this important effect
of reheating, while simpler approximations such as setting $T_{-}=T_{+}$ in
Eq.~(\ref{Fdr}) (see e.g. \cite{ekns10}) \emph{overestimate} the driving force.

\subsection{Friction force}

As already discussed, the friction must be obtained from microphysics
considerations which are much more involved than the calculation of
$F_{\mathrm{dr}}$. Here we shall use instead the phenomenological damping
(\ref{Kfeno}). Hence, we have
\begin{equation}
F_{\mathrm{fr}}=\int_{-}^{+}\frac{f(\phi )\,u^{\mu }\partial _{\mu }\phi }{\sqrt{1+[g(\phi
)\,u^{\mu }\partial _{\mu }\phi ]^{2}}}\,\partial_z\phi \,dz . \label{Ffrdef}
\end{equation}
For the field profile (\ref{fieldprof}), we have
\begin{equation}
u^\mu\partial_\mu\phi=\gamma(\partial_0\phi+v\partial_z\phi) =
\phi_0'\gamma[\gamma_w(v-v_w) + \dot\gamma_w(z-z_w)]. \label{udfi}
\end{equation}
The term $\dot\gamma_w(z-z_w)$ vanishes in the stationary case. In the runaway
regime we can also neglect it, since $\phi_0'$ vanishes out of the thin
interface\footnote{More precisely, $\dot\gamma_w$ is proportional to the proper
acceleration $\gamma_w^3\dot v_w$ which, according to Eq.~(\ref{eqwall}), is
bounded by $\sim F_{\mathrm{dr}}/\sigma_0$, while $z-z_w$ is bounded by
$l_0/\gamma_w$, where $l_0$ is the wall width at rest.}. We thus have
\begin{equation}
F_{\mathrm{fr}}=\int_{-}^{+}\frac{\gamma_w\gamma(v-v_w)
f(\phi_0)(\phi_0')^2}{\sqrt{1+\gamma_w^2\gamma^2(v-v_w)^2 g^2(\phi_0) (\phi_0')^2}} \,dz ,
\label{Ffrfenpl}
\end{equation}
where we have used the change of variable of integration $\gamma_w
(z-z_w)\to z$. It is easy to see that in the limit $\gamma_w\to \infty$ we
have $F_{\mathrm{fr}}\sim$ constant, while for $v_w\ll 1$ (which implies
$v\ll 1$ as well) we have $F_{\mathrm{fr}}\sim v_w$. The result
(\ref{Ffrfenpl}) is equivalent to that of Ref.~\cite{ariel13}, as can be
seen from the transformation $\gamma v\to \gamma\gamma_w(v-v_w)$ from the
wall frame to the plasma frame.

The integral in Eq.~(\ref{Ffrfenpl}) is of the form $\int
[\phi_0'(z)]^2F(z)\,dz$, where the  function $[\phi_0'(z)]^2$ vanishes outside
the wall and peaks at the center of the latter. Therefore, we can write this
integral as $F(\bar z)\int [\phi_0'(z)]^2dz=F(\bar z)\sigma_0$, where $\bar z$
is some point near the wall center. We thus obtain
\begin{equation}
F_{\mathrm{fr}}=
\frac{\eta\gamma_w\bar\gamma(\bar v-v_w)}{\sqrt{1+
\lambda^2\gamma_w^2\bar\gamma^2(\bar v-v_w)^2}}, \label{Ffrfeno}
\end{equation}
where  $\bar\gamma,\bar v$ are the values of $\gamma,v$ at the center of the
wall, $\eta=\sigma_0f(\phi_0(\bar z))$, and $\lambda$ is similarly given by the
function $g$ and details of the wall profile. We shall regard $\eta$ and
$\lambda$ as free parameters which can be chosen appropriately to give the
correct numerical values of the friction in the NR and UR limits. On the other
hand, we shall approximate the value of $\bar v$ by the average\footnote{In
Ref.~\cite{ariel13}, a different approximation was used, in which the whole
function of $\bar v$ in Eq.~(\ref{Ffrfeno}) was replaced by its average value.
We have checked that there is no significant numerical difference.}
\begin{equation}
\bar v=({v_-+v_+})/{2}=v_-/2, \label{vmedia}
\end{equation}
and $\bar\gamma=1/\sqrt{1-\bar v^2}$.

For non-relativistic velocities, Eq.~(\ref{Ffrfeno}) gives a friction force
which is proportional to the relative velocity, $F_{\mathrm{fr} }=-\eta
\,(v_{w}-\bar v)$, as expected. For a specific model, the value of $\eta $ can
be obtained by comparison with the result of a non-relativistic microphysics
calculation. Therefore, we use the notation $\eta= \eta _{NR}$. It is out of
the scope of the present work to compute the friction for specific models.
General approximations for $\eta _{NR}$  as a function of the parameters for a
variety of models can be found in Ref. \cite{ms10}. The friction coefficient
depends on temperature. In particular, it decreases as $T$ decreases, since the
friction depends on the particle populations. Nevertheless, in contrast to
$F_{\mathrm{dr}}$, the friction is not sensitive to the temperature difference
$T_{c}-T$. Therefore, it is not very sensitive to hydrodynamics. For specific
calculations, in this work we shall assume  that, roughly, $\eta_{NR}\sim
T_{+}^{4}$.

In the ultra-relativistic case $\gamma_w\gg 1$, Eq.~(\ref{Ffrfeno}) gives
$F_{\mathrm{fr}}= -\eta/\lambda$. Therefore, we define the UR friction
coefficient $\eta _{UR}=-\eta/\lambda$, so that  $F_{\mathrm{fr}}= -\eta
_{UR}v_w$. The value of this parameter for a specific model can be obtained
from the microphysics result (\ref{Fnet}). This result, however, gives the
total UR force $F_{\mathrm{net}}$ rather than the friction. Notice that the
last term in Eq.~(\ref{Fnet}) includes the finite-temperature part of the
driving force, $F_{\mathrm{hyd}}$, as well as the friction. We may obtain the
UR friction force as $F_{\mathrm{fr}}=F_{\mathrm{net}}-F_{ \mathrm{dr}}$,
taking into account the UR limit of the driving force. The latter is given by
Eqs.~(\ref{Fdrbag}), (\ref{tmetmabag}), and (\ref{tmevmerun}). We obtain
\begin{equation} \frac{\eta_{UR}}{aT_+^4}=
\alpha-\bar{F}-\alpha_{c}
\sqrt{\frac{1+3(\alpha-\bar{F})}{1-3\alpha_{c}}} .  \label{etaur}
\end{equation}
For a given model, the quantities $\alpha=L/(3w_+)$ and $\bar
F=4F_{\mathrm{net}}/(3w_+)$ can be obtained as functions of the nucleation
temperature $T_+$. We remark that, although decomposing the total force into
driving and friction forces is not relevant for the runaway regime, determining
the UR value of the friction component is relevant for a correct use of
Eq.~(\ref{Ffrfeno}) as an interpolation between the NR and UR cases. In terms
of the friction coefficients $\eta_{NR},\eta_{UR}$, we have
\begin{equation}
F_{\mathrm{fr}}=-\frac{\eta_{NR}\eta_{UR}\,
\gamma_w\bar\gamma(v_w-\bar v)}{\sqrt{\eta_{UR}^2+
\eta_{NR}^2\,\gamma_w^2\bar\gamma^2(v_w-\bar v)^2}}\label{Ffrfeneta}.
\end{equation}

It is worth commenting on previous approaches.  A similar, but simpler,
phenomenological model for the friction  was considered in Ref.~\cite{ekns10}.
The approximation involves a single free parameter and is equivalent to setting
$\lambda=1$ in Eq.~(\ref{Ffrfeno}). Although this model gives a friction which
saturates at high $\gamma_w$, it is numerically incorrect as it corresponds to
the case $\eta _{UR}=\eta _{NR}$ (besides, the approximation $T_-=T_+$ was used
in \cite{ekns10} for the driving force; we shall discuss on this approximation
below). The phenomenological model (\ref{Kfeno}) was already considered in
Ref.~\cite{ariel13}. The resulting friction is equivalent to
Eq.~(\ref{Ffrfeno}). However, in \cite{ariel13} the hydrodynamics was neglected
for runaway walls (but not for stationary solutions). That is, the relation
$T_{-}=T_{+}$ was assumed for the runaway case. This results in a different
value of  $\eta_{UR}$, as the right-hand side of Eq.~(\ref{etaur}) becomes
$\alpha-\bar F-\alpha_c$. Since, on the other hand, the hydrodynamics was taken
into account for detonations, the stationary solutions did not match
continuously the accelerated ones.  This is not correct since, as we have seen,
the hydrodynamics of a runaway wall is similar to that of the detonation, and
matches the latter for $F_{\mathrm{net}}=0$.

\subsection{The wall velocity}

From Eqs.~(\ref{eqforces}), (\ref{Fdrbag}) and (\ref{Ffrfeneta}), we have, for
detonations or runaway walls,
\begin{equation}
\frac{F_{\mathrm{net}}}{aT_+^4}=
\alpha-\alpha_c\frac{T_{-}^{2}}{T_{+}^{2}}
-\frac{\bar\eta_{NR}\bar\eta_{UR}\,(v_w-\bar v_-)}{\sqrt{\bar\eta_{UR}^2(1-v_w^2)
(1-\bar v_-^2)+
\bar\eta_{NR}^2\,(v_w-\bar v_-)^2}} \label{eqwallbag}
\end{equation}
where $\bar v=v_-/2$, and we use the notation $\bar\eta=\eta/(aT_+^4)$ for the
two friction coefficients. In the UR limit, Eq.~(\ref{eqwallbag}) becomes
\begin{equation}
\bar F=\alpha-{\alpha_c}{T_{-}^{2}}/{T_{+}^{2}}
-\bar\eta_{UR}. \label{eqwallbagur}
\end{equation}
We remark again that the latter equation is just a decomposition of the UR net
force, which actually defines the value of the friction coefficient
$\eta_{UR}$, while the former gives an equation of motion for the wall away
from that limit. In particular, if the microphysics computation of the net
force, Eq.~(\ref{Fnet}), gives $F_{\mathrm{net}}<0$, it means that, in fact,
the wall will never reach the UR regime. Nevertheless, the UR calculation is
still useful and Eq.~(\ref{eqwallbagur}) makes sense. The interpretation is
that the UR friction is so high that the driving force cannot compensate it. In
this case we just obtain $\bar\eta_{UR}$ from Eq.~(\ref{etaur}), and then
compute the steady-state wall velocity by setting $F_{\mathrm{net}}=0$ in
Eq.~(\ref{eqwallbag}).

To solve for $v_w$, we must use Eqs.~(\ref{vme}-\ref{tmetmabag}) for $v_-$ and
$T_-$. It is worth mentioning that, for detonations, the result does not depend
on the wall being spherical or planar, since all the quantities appearing in
Eq.~(\ref{eqwallbag}) are the same in the two cases. This is because the
relations between $v_-,T_-$ and $v_+,T_+$ are the same for spherical or planar
walls. Besides, for detonations  the conditions in front of the wall (i.e.,
$v_+,T_+$) are also the same (in contrast, for deflagrations, the fluid in
front of the wall is perturbed differently for planar or spherical walls).

\begin{figure}[bth]
\centering
\epsfysize=6cm \leavevmode \epsfbox{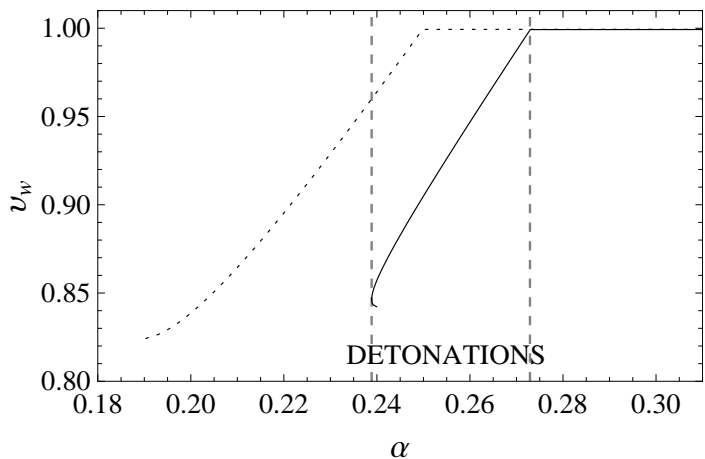}
\caption{The wall velocity
for $\alpha_c=0.05$, $ \bar\eta _{UR}=0.2$, and $\protect\eta
_{NR}=\protect\eta _{UR}$. The dotted line corresponds to the approximation $T_-=T_+$.}
\label{figfric}
\end{figure}

We show the result in Fig.~\ref{figfric} (solid line) for fixed values of the
friction parameters and varying the bag quantity $\alpha$. For concreteness,
and in order to compare with previous results, we consider the case $\eta
_{UR}=\eta _{NR}$ (for other cases and different parameter variations, see
\cite{ariel13}). The vertical dashed lines delimit the weak detonation
solutions. Increasing $\alpha$ generally increases the driving force and,
consequently, the wall velocity. The figure does not show the deflagration
cases, for which $v_w\lesssim c_-$ (there is a discontinuity between
deflagrations and detonations).

The dotted line in Fig.~\ref{figfric} is obtained by neglecting  the reheating
in the calculation of the driving force, i.e., setting $T_{-}=T_{+}$, for which
the driving-force term in Eq.~(\ref{eqwallbag}) becomes $\alpha-\alpha_c$. This
was used as an approximation in Ref.~\cite{ekns10}. We consider it here in
order to appreciate the role of hydrodynamics. Quantitatively, we see that this
approximation overestimates the driving force, as we obtain higher values of
the velocity. Besides, we observe a significant qualitative difference between
the two results at the lower end of the detonation curve. This end corresponds
to the Jouguet point. Since the hydrodynamics becomes very strong near this
point, Eq.~(\ref{eqwallbag}) gives two solutions for $v_{w}$, while the
approximation $ T_{-}=T_{+}$ completely misses this effect. In
Ref.~\cite{mm14deto} it was shown that weak detonations corresponding to the
lower branch of solutions are unstable.

The value of $\alpha$ for which the detonation reaches the ultra-relativistic
regime in Fig.~\ref{figfric} can be obtained from Eq.~(\ref{eqwallbagur})
which, for $\bar F=0$, gives  $\alpha
={\alpha_c}{T_{-}^{2}}/{T_{+}^{2}}+{\bar\eta_{UR}}$. Notice that $T_-$ actually
depends on $\alpha$ through Eqs.~(\ref{tmetmabag},\ref{tmevmeurdeto}),
${T_{-}^{4}}/{T_{+}^{4}}=({1+3\alpha})/({1-3\alpha_{c}})$. Thus, for given
values of $\alpha_c$ and $\bar\eta_{UR}$ we have a quadratic equation for
$\alpha$ (or for $T_-$), which yields
\begin{equation}
\alpha= \alpha_{c}\,{T_{0}^{2}}/{T_{+}^{2}}+{\bar\eta _{UR}}\equiv\alpha_0,
\label{L0deeta}
\end{equation}
where $T_0$ is the corresponding value of $T_-$, given by
\begin{equation}
\frac{T_{0}^{4}}{T_{+}^{4}}=\frac{3\alpha_{c}}{2(1-3\alpha_{c})}
\left[1+\sqrt{ 1+\frac{4(1+3\bar\eta _{UR})(1-3\alpha_c)}{
3\alpha_c^2}}\right]. \label{tmedeeta}
\end{equation}
For $\alpha>\alpha_0$, the steady-state equation gives $v_w>1$, which actually
indicates that the friction force cannot compensate the driving force and we
have $F_{\mathrm{net}}>0$, i.e., a runaway wall.

\begin{figure}[bth]
\centering
\epsfysize=6cm \leavevmode \epsfbox{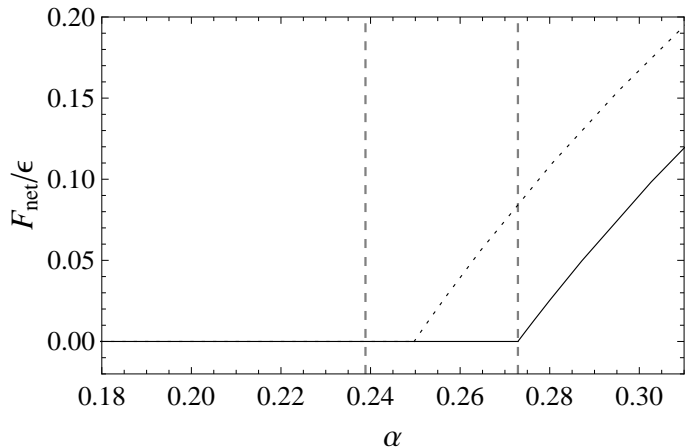}
\caption{The net force corresponding to Fig.~\ref{figfric}.}
\label{figfricF}
\end{figure}

In the runaway regime, we have  a proper acceleration which is proportional to
the net force. It is interesting to calculate the value of  $F_{\mathrm{net}}$
corresponding to Fig.~\ref{figfric}, which can be obtained from
Eq.~(\ref{eqwallbagur}) [although for a given model one would rather compute
$F_{\mathrm{net}}$ directly from Eq.~(\ref{Fnet}), and then determine
$\eta_{UR}$]. We must take into account the dependence of $T_{-}$ on
$F_{\mathrm{net}}$, which is given by Eqs.~(\ref{tmetmabag},\ref{tmevmerun}),
\begin{equation}
T_-^4/T_+^4=(1+3\alpha-3\bar F)/(1-3\alpha_c). \label{tmeur}
\end{equation}
From Eqs.~(\ref{tmeur}) and (\ref{eqwallbagur}) we obtain quadratic equations
for $\bar F$ and $T_-/T_+$ as functions of $\alpha$, $\alpha_c$, and
$\bar\eta_{UR}$. Nevertheless, the dependence on $\alpha$ cancels in the
equation for $T_-$, and we obtain
\begin{equation}
T_-=T_0,\quad
\bar{F}=\alpha-\alpha_cT_0^2/T_+^2-\bar\eta_{UR}=\alpha-\alpha_{0},
\label{Fdeeta}
\end{equation}
with $\alpha_0$ and $T_0$ given by Eqs.~(\ref{L0deeta}-\ref{tmedeeta}). On the
other hand, if we use the approximation $T_-=T_+$, we just neglect
Eq.~(\ref{tmeur}), while Eq.~(\ref{eqwallbagur}) gives $\bar F=\alpha-\alpha_c
-\bar\eta_{UR}$. This result can also be written in the form $\bar
F=\alpha-\alpha_{0}$, but the value of $\alpha_0$ is different, namely,
$\alpha_0=\alpha_c+\bar\eta_{UR}$. In Fig.~\ref{figfricF} we plot the net
force, normalized to its maximum value $\varepsilon$, as a function of $\alpha$
for the parameters of Fig.~\ref{figfric}. We see that neglecting the
hydrodynamics gives a higher wall acceleration.

Comparing the solid lines of Figs.~\ref{figfric} and \ref{figfricF}, we see
that the detonation solution matches the runaway solution at $\alpha=\alpha_0$.
In Ref.~\cite{ariel13} this matching does not occur,  due to the assumption of
different hydrodynamics. As a consequence, the two kinds of solutions were
found to coexist in a small parameter range. We do not find such a coexistence
of detonation and runaway solutions here, since the hydrodynamics is continuous
with $v_w$ and $\bar F$. In fact, coexistence of solutions could arise also due
to strong hydrodynamics, even if the hydrodynamics varies continuously with the
parameters. For instance, we have multiple weak-detonation solutions near the
Jouguet point, even though $v_-,T_-$ are continuous functions of $\alpha,v_w$.
This does not happen in the UR limit, since the perturbations of the fluid vary
continuously with $\alpha,v_w$, and $\bar F$ and, besides, the hydrodynamics is
weaker.

\subsection{Microphysics and  released energy}

In Sec.~\ref{hidro} we computed the fractions of the released vacuum energy
which go into bulk motions of the fluid and into kinetic energy of the bubble
wall, as functions of the quantities $v_w$, $\bar F=(4/3)F_{\mathrm{net}}/w_+$,
and $\alpha=L/(3w_+)$. For a given model, the nucleation temperature and the
thermodynamical quantities $L,w_+$ can be calculated from the
finite-temperature effective potential (\ref{ftot}), and the net force can be
readily computed from Eq.~(\ref{Fnet}). This gives the values of $\alpha$ and
$\bar F$. The steady-state wall velocity can be obtained from
Eq.~(\ref{eqwallbag}), after determining the friction coefficients  $\eta_{UR}$
and $\eta_{NR}$. The value of $\eta_{UR}$ can be obtained from $\bar F$ using
Eq.~(\ref{etaur}), while $\eta_{NR}$ must be obtained from a microphysics
calculation. Such a computation is beyond the scope of this paper. Here, we
shall only consider the energy distribution among the fluid and the wall for
the parameter variation of Figs.~\ref{figfric} and \ref{figfricF}. As already
discussed, this parameter variation becomes rather artificial in the runaway
regime. It is useful, though,  for a comparison with previous results.

\begin{figure}[bt]
\centering
\epsfxsize=15cm \leavevmode \epsfbox{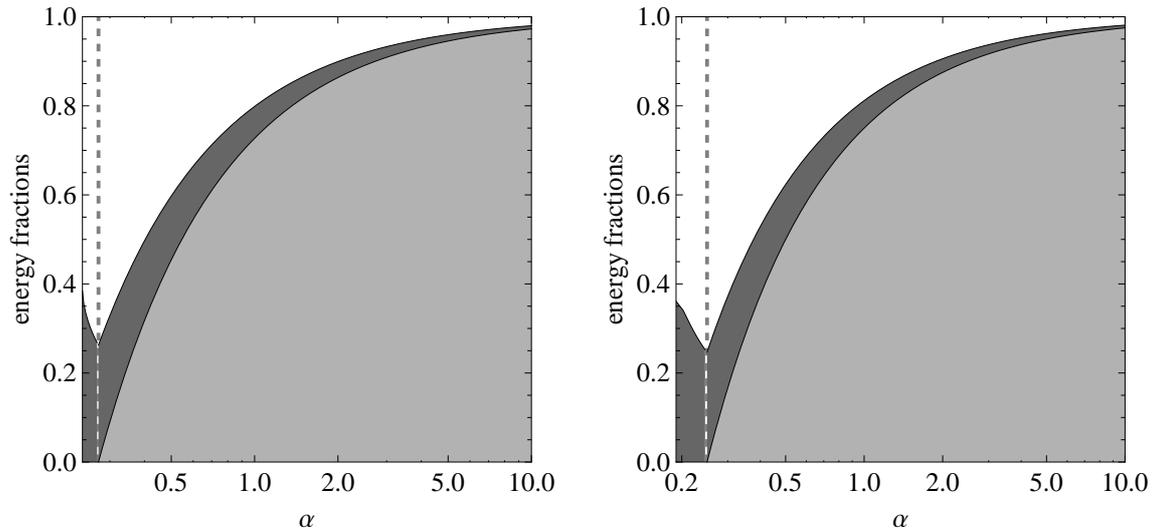}
\caption{The fractions of vacuum energy which go to the wall and the fluid,
for the same case of Figs.~\ref{figfric} and \ref{figfricF}.
The right panel corresponds to neglecting the reheating in the calculation
of the driving force.}
\label{figbudget}
\end{figure}

The fraction of energy accumulated in the interface, $\kappa_{\mathrm{wall}}$,
is just given by $F_{\mathrm{net}}/\varepsilon$, which is  plotted in
Fig.~\ref{figfricF}. In Fig.~\ref{figbudget} we consider a wider range of
runaway solutions, and we plot separately (in the right panel) the result
obtained by using the approximation $T_{-}=T_{+}$ in the calculation of the
driving force. The value of $\kappa_{\mathrm{wall}}$ is represented by the
height of the light shade. Thus, the curves delimiting this region in the left
and right panels of Fig.~\ref{figbudget} correspond, respectively, to the solid
and dotted lines of Fig.~\ref{figfricF}. Following Ref.~\cite{ekns10}, we plot
the value of $\kappa _{\mathrm{fl}}$ (for spherical bubbles) added to that of
$\kappa_{\mathrm{wall}}$. This gives the upper curves delimiting the dark shade
regions. Hence, the fraction of $\varepsilon$ which goes into bulk fluid
motions is represented by the dark shade. Accordingly, the white region
indicates the portion of $\varepsilon $ which goes into reheating\footnote{In
fact, thermal energy is released as well as vacuum energy (whence
$L>\varepsilon$). As a consequence, the  white regions in Fig.~\ref{figbudget}
only represent a part of the total energy which goes into reheating of the
plasma. For a more detailed discussion, see \cite{lm15}.}. The vertical line
separates the detonation and runaway regimes. We include the complete
detonation range (which is different in the two panels).

The right panel of Fig.~\ref{figbudget} agrees with the results of
Ref.~\cite{ekns10}. The values of the parameters, namely, $\eta _{NR}=\eta
_{UR}=0.2$, $a_{-}/a_{+}=1-3\alpha_{c}=0.85$, correspond to one of the cases
considered in that work (cf.~the left panel of Fig.~12 in \cite{ekns10}). We
observe that the two panels of Fig.~\ref{figbudget} are qualitatively similar,
particularly for the runaway regime, where the hydrodynamics is weaker. The
difference is more apparent for detonations, where the hydrodynamics is
strongest. In fact, the Jouguet point is never reached in the left panel. The
quantitative difference between the two calculations is better appreciated in
Figs.~\ref{figfric} and \ref{figfricF}. For a given  $\alpha$, neglecting the
hydrodynamics gives higher wall velocities and accelerations and, hence, a
larger $\kappa_{\mathrm{wall}}$ and a smaller $\kappa_{\mathrm{fl}}$. It is
worth emphasizing that in both panels we have used the results from
Sec.~\ref{hidro} in terms of $\alpha$, $v_w$ and $\bar F$, and the discrepancy
originates in the computation of $v_w$ and $\bar F$.

For the runaway case we may obtain simple semi-analytical expressions for the
efficiency factors as functions of the quantities $\alpha$, $\alpha_c$, and
$\bar\eta_{UR}$. From Eqs.~(\ref{enwall}) and (\ref{Fdeeta}), we have
\begin{equation}
\kappa_{\mathrm{wall}}=1-{\alpha_{0}}/{\alpha}.  \label{ewallfric}
\end{equation}
This is valid for the two plots of Fig.~\ref{figbudget}, with different values
of $\alpha_0(\alpha_c,\bar\eta_{UR})$. Since $\kappa_{\mathrm{wall}}$ gives the
fraction of $\varepsilon$ which goes to kinetic energy of the wall,
Eq.~(\ref{ewallfric}) indicates that a fraction $\alpha_{0}/\alpha$ goes to the
fluid (either to bulk motions or reheating). Moreover, using Eq.~(\ref{Fdeeta})
in Eqs.~(\ref{kapparun}) and (\ref{tmevmerun}), we have
\begin{equation}
\kappa _{\mathrm{fl}}^{\mathrm{run}}=
\frac{4(1+3\alpha_{0})I_{1}(v_-)}{\alpha}, \quad \mbox{with}\quad
v_-=\frac{3\alpha_{0}}{2+3\alpha_{0}}.
\end{equation}
Hence, the runaway efficiency factor can be written as
\begin{equation}
\kappa _{\mathrm{fl}}^{\mathrm{run}}=\kappa _{UR}^{\mathrm{det }}\,{\alpha_{0}}/{\alpha},
\label{krunfric}
\end{equation}
where $\kappa _{UR}^{\mathrm{det}}\equiv\kappa
_{\mathrm{fl}}^{\mathrm{det}}(\alpha_0,v_w=1)$ is the efficiency factor of the
UR detonation. The functional dependence of Eqs.~(\ref{ewallfric}) and
(\ref{krunfric}) on $\alpha$ agrees with the results of Ref. \cite{ekns10}. The
quantitative difference, which is illustrated by the two panels of
Fig.~\ref{figbudget}, is due to different values of $\alpha_0$ and $\kappa
_{UR}^{\mathrm{det }}$ (in the notation of \cite{ekns10}, $ \alpha_0=\alpha
_{\infty}$ and $\kappa _{UR}^{\mathrm{det }}=\kappa _{\infty }$). As already
discussed, this discrepancy is due essentially to a different treatment of
hydrodynamics.

In Ref.~\cite{ekns10} an expression for the quantity $\alpha_0$ is provided in
terms of microphysics parameters. We may obtain a similar expression as
follows. If we identify the difference $V(\phi_+)-V(\phi_-)$ in
Eq.~(\ref{Fnet}) with the bag constant $\varepsilon$ (notice, though, that the
minima, particularly $\phi_-$, are temperature dependent), then the UR net
force vanishes for
\begin{equation}
\varepsilon_0=\sum_i g_i c_i{T_+^2m_i^2(\phi_-)}/{24}. \label{eps0}
\end{equation}
For a given $T_+$, this is the value of $\varepsilon$ corresponding to
$\alpha_0$. Thus, we have $\alpha_0=4\varepsilon_0/(3w_+)$, which can be used
in Eqs.~(\ref{ewallfric}-\ref{krunfric}). We remark that this approach involves
more approximations than those used in Sec.~\ref{hidro}, where we obtained
$\kappa_{\mathrm{fl}} ^{\mathrm{run}}$ directly as a function of $\bar F$.

\section{Gravitational waves} \label{gw}

The efficiency factors $\kappa_{\mathrm{fl}}$ and $\kappa_{\mathrm{wall}}$ give
the fractions of the released energy which go into fluid motions and into the
wall, respectively. The values of these factors are the key quantities in the
different mechanisms of gravitational wave (GW) generation in a first-order
phase transition. Three mechanisms of GW generation have been considered in the
literature, namely, bubble collisions, turbulence, and sound waves. The bubble
collisions mechanism assumes that the energy-momentum tensor which sources the
GWs is concentrated in thin spherical shells \cite{gwcol,hk08}. For
detonations, this is not a bad approximation during the phase transition, since
a portion $\kappa_{\mathrm{fl}}\varepsilon\Delta V_b$ of the released vacuum
energy $\varepsilon\Delta V_b$  is concentrated as kinetic energy of the fluid
in a region which follows the bubble wall supersonically.  This is also a good
approximation for runaway walls, for which another portion
$\kappa_{\mathrm{wall}}\varepsilon\Delta V_b$ of the  vacuum energy is
accumulated in the infinitely thin interface. Hence, the total energy involved
in this mechanism is proportional to
$\kappa_{\mathrm{tot}}=\kappa_{\mathrm{wall}}+ \kappa_{\mathrm{fl}}$ (with
$\kappa_{\mathrm{wall}}=0$ in the detonation case).

On the other hand, fluid motions can remain long after the completion of the
phase transition and continue producing GWs. This may happen by two mechanisms,
namely, magnetohydrodynamic (mhd) turbulence \cite{gwturb} or sound waves
\cite{gwsound}. Since these are long-lasting sources, they are generally more
efficient than bubble collisions. However, the energy involved in these ``fluid
motions'' mechanisms is proportional to $\kappa_{\mathrm{fl}}$ alone. In the
runaway regime the hydrodynamics becomes weaker and the energy in the fluid is
suppressed. As a consequence, it is not clear a priori whether these mechanisms
will still dominate over bubble collisions.

Depending on the generation mechanism, the peak frequency of the GW spectrum is
determined by a characteristic time or a characteristic length of the source.
For a first-order phase transition, the time scale is given by its duration
$\Delta t$, while the length scale is given by the average bubble radius $R\sim
v_w\Delta t$. For detonations or runaway walls, we have $R\sim\Delta t$.
Therefore, the characteristic frequency at the formation of GWs is given by
$f_{p*}\sim 1/\Delta t$. The corresponding frequency today (after redshifting)
would be given by
\begin{equation}
f_p\sim  10^{-5}\,\mathrm{Hz}\left( \frac{g_{\ast }
}{100}\right) ^{1/6}\left( \frac{T}{100\,\mathrm{GeV}}\right)  \frac{1 }{H\Delta t},
\end{equation}
where $H$ is the Hubble rate during the phase transition, $g_*$ is the number
of relativistic degrees of freedom, and $T$ is the temperature at which the
phase transition occurred, namely, $T\approx T_+\lesssim T_c$.

It is interesting to consider  the electroweak phase transition, for which we
have $T_c\simeq 100$GeV and  $g_*\simeq 100$. The duration of the phase
transition may vary from $\Delta t\sim 10^{-5}H^{-1}$ for very weak phase
transitions  to $ \Delta t\sim H^{-1}$ for very strong phase transitions. For
the sake of concreteness, we shall consider $H\Delta t =10^{-1}$, corresponding
to strong phase transitions, which is consistent with having detonations or
runaway walls. This gives $f_p\sim 0.1\mathrm{mHz}$, which is close to the peak
sensitivity of the planned space-based observatory eLISA \cite{elisa}. It is
customary to express the energy density of gravitational radiation in terms of
the quantity
\begin{equation}
h^2\Omega _{GW}\left( f\right) =\frac{h^2}{\rho _{c}}\frac{d\rho
_{GW}}{d\log f},
\end{equation}
where $\rho _{GW}$ is the energy density of the GWs, $f$ is the frequency, and
$\rho _{c} $ is the critical energy density today, $\rho _{c}=3H_{0}^{2}/8\pi
G$, with $H_0=100\, h\, \mathrm{ km \, s}^{-1} \mathrm{Mpc}^{-1}$, and
$h=0.72$. The peak sensitivity of eLISA may be in the range $\Omega_{GW}\sim
10^{-14}-10^{-10}$, depending on its final configuration.

An approximation for the spectrum of GWs from bubble collisions was given in
Ref.~\cite{hk08}. For the peak amplitude that would be observed today we have
\begin{equation}
h^{2}\Omega _{\mathrm{\mathrm{coll}}}=1.67\times 10^{-5}\left(
\frac{\kappa_{\mathrm{tot}} \alpha}{1+\alpha}
\right) ^{2}\left( \frac{100}{g_{\ast
}}\right)^{1/3} \left(\frac{0.11v_{w}^{3}}{
0.42+v_{w}^{2}}\right) \left(\Delta t H\right)^{2}.  \label{omcol}
\end{equation}
The spectrum from mhd turbulence was calculated using  analytic approximations
in Ref.~\cite{cds09}. For the peak amplitude we have \cite{cds09,cds10}
\begin{equation}
h^{2}\Omega _{\mathrm{turb}}=2.6\times 10^{-5}
\left( \frac{\kappa_{\mathrm{fl}} \alpha}{1+\alpha}\right) ^{3/2}
\left( \frac{100}{g_{\ast }}\right) ^{1/3}\frac{{v_w\Delta t H}
}{1+4\pi 3.5/(v_w\Delta tH) }.
\label{omturb}
\end{equation}
Regarding the GW spectrum from sound waves, a fit to the numerical results of
Ref.~\cite{gwsound} was given in \cite{elisasci}. For the peak intensity we
have
\begin{equation}
h^{2}\Omega _{sw}=2.65\times 10^{-6}\left(\frac{\kappa_{\mathrm{fl}}
\alpha}{1+\alpha}\right) ^{2}\left( \frac{100}{g_{\ast
}}\right) ^{1/3}{v_w\Delta t H}.  \label{omsw}
\end{equation}
The quantity $\alpha=\varepsilon/(aT_+^4)$ gives the ratio of the vacuum energy
density to the radiation energy density. For the electroweak phase transition
we generally have $\alpha<1$, i.e., radiation dominates. Hence, for $g_*\sim
100$, $v_w\sim 1$, and $\Delta t H\sim 10^{-1}$ we have
$h^{2}\Omega_\mathrm{coll}\sim 10^{-8}(\kappa_{\mathrm{tot}}\alpha)^2$,
$h^{2}\Omega_\mathrm{turb}\sim 10^{-8}(\kappa_{\mathrm{fl}}\alpha)^{3/2}$, and
$h^{2}\Omega_\mathrm{sw}\sim 10^{-7}(\kappa_{\mathrm{fl}}\alpha)^{2}$. We see
that the numerical values are similar, and the precise results will depend on
details of the phase transition dynamics. In particular, the values of $\alpha$
and the efficiency factors will determine which of these sources is dominant.

\begin{figure}[bt]
\centering
\epsfxsize=15cm \leavevmode \epsfbox{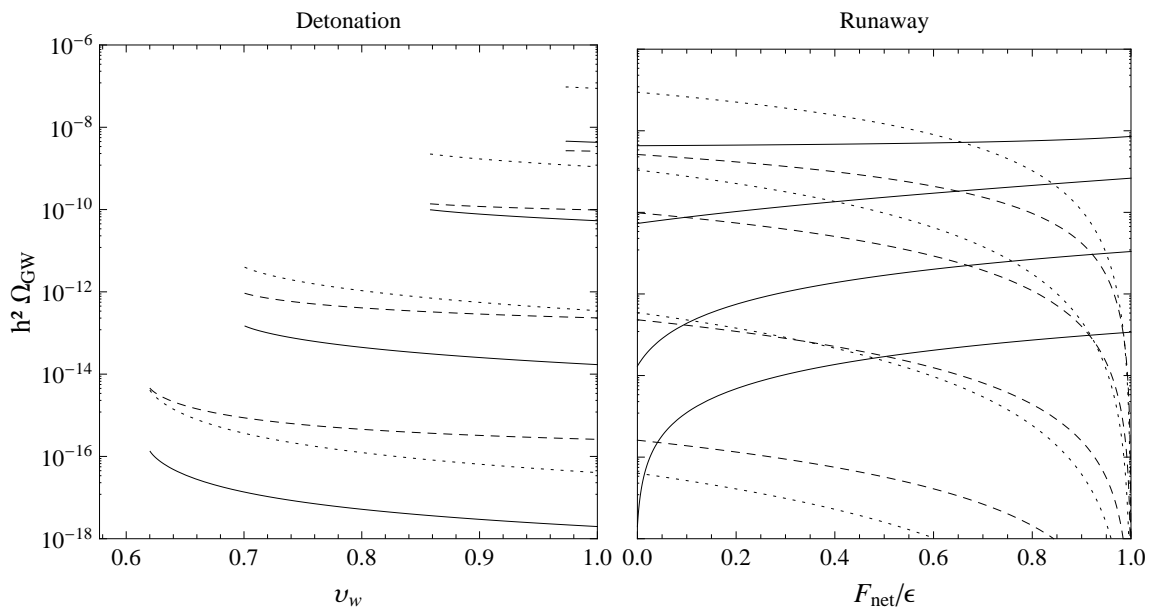}
\caption{The peak amplitude of the gravitational wave spectrum from bubble collisions
(solid lines), turbulence (dashed lines), and sound waves (dotted lines), for $g_*=100$,
$\Delta t=0.1$, and for a few values of $\alpha$. From bottom to top, we have
$\alpha=0.003$, $0.03$, $0.3$, and $3$.}
\label{figgw}
\end{figure}

The efficiency factors depend on $\alpha$. Besides, they depend on the wall
velocity (in the detonation case) and on the net force (in the runaway case).
In Fig.~\ref{figgw} we plot the GW intensities (\ref{omcol}-\ref{omsw}) as
functions of the wall velocity and the net force, for  some  values of
$\alpha$.
We see that the different sources dominate in different parameter regions.
However, in the case of stationary walls the GW signal from bubble collisions
is generally smaller, as expected. Besides, in the detonation case all the
curves behave similarly as functions of $v_w$. This is due to the dependence of
the GW amplitudes  on $\kappa_\mathrm{fl}$, which decreases with $v_w$
(cf.~Figs.~\ref{figplesf} and \ref{figkesf}). In contrast, for runaway walls,
the GW signal from bubble collisions grows with the net force, while the other
signals decrease. This is because the former depends on
$\kappa_\mathrm{wall}=F_{\mathrm{net}}/\varepsilon$, as already discussed.
Also, as expected, all the signals grow with $\alpha$ for a given value of
$v_w$ or $F_{\mathrm{net}}/\varepsilon$.

\begin{figure}[bt]
\centering
\epsfysize=7cm \leavevmode \epsfbox{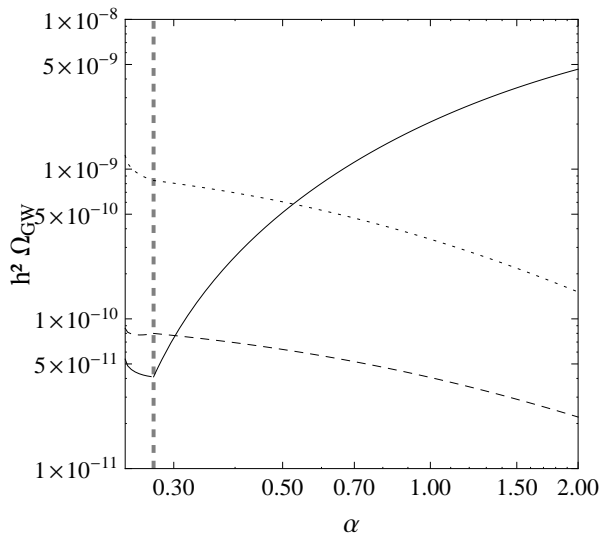}
\caption{The peak amplitude of the gravitational wave spectrum from bubble collisions
(solid line), turbulence (dashed line), and sound waves (dotted line) as functions of
$\alpha$, for $g_*=100$,
$\Delta t=0.1$, and the rest of the parameters as in Figs.~\ref{figfric}-\ref{figbudget}.
The vertical line separates detonations from runaway solutions.}
\label{figgwfric}
\end{figure}

The quantities  $v_w,F_{\mathrm{net}}/\varepsilon$ and $\alpha$ are not
actually independent. As we have seen, for fixed values of the friction
parameters, $v_w$ and $F_{\mathrm{net}}/\varepsilon$ are increasing functions
of $\alpha$. In such a case, $\kappa_\mathrm{fl}$ actually decreases with
$\alpha$, as shown in Fig.~\ref{figbudget}, while $\kappa_\mathrm{wall}$
increases. In Fig.~\ref{figgwfric} we plot the GW amplitudes corresponding to
that parameter variation.
We see that the decrease of  $\kappa_\mathrm{fl}(\alpha)$ is reflected in  the
GW amplitudes, even though the latter depend on the product
$\kappa_\mathrm{fl}(\alpha)\, \alpha$. Indeed, the signals from fluid motions
generally decrease with $\alpha$, while the signal from bubble wall collisions
grows in the runaway regime due to the increase of $\kappa_{\mathrm{wall}}$.

It is important to notice that this behavior of the GW signals with the
quantity $\alpha$ was obtained by fixing several parameters, such as the
friction coefficients $\eta_{NR},\eta_{UR}$, the bag parameter
$\alpha_c=\varepsilon/(aT_c^4)$ (which is equivalent to fixing $a_-/a_+$), as
well as the time scale $\Delta t$. In a concrete model, all these quantities
will vary together with $\alpha$, as all of them depend on the model
parameters. We shall consider concrete models elsewhere. In any case, values of
$\alpha$ in the range $0.3\lesssim \alpha\lesssim 1$ are possible in a very
strong electroweak phase transition. Hence, Fig.~\ref{figgwfric} shows that the
GWs generated by any of the mechanisms at this phase transition  may be
observable by eLISA.

\section{Conclusions} \label{conclu}

Several hydrodynamic modes are possible for the growth of a bubble in a
cosmological first-order phase transition. These include steady-state walls
propagating as deflagrations or detonations \cite{hidro},  accelerated
(runaway) walls \cite{bm09}, or even turbulent motion associated with wall
corrugation \cite{inst}. Which of these propagation modes will a phase
transition front take, depends on several factors, such as the amount of
supercooling and the friction of the bubble wall with the plasma. In this work
we have studied the fastest of these modes, namely, ultra-relativistic
detonations and runaway solutions, which may give an important gravitational
wave signal from the phase transition.

The generation of gravitational waves depends on the released energy, which is
usually measured by the ratio $\alpha$ of the vacuum energy density to the
radiation energy density. It is also important how this energy is distributed,
as the wall moves, among the bubble wall and the fluid. For a steady-state
wall, all the released energy goes to the fluid, either to reheating or to bulk
motions. The fraction $\kappa_{\mathrm{fl}}$ of the released vacuum energy
which goes into bulk motions is relevant for the formation of gravitational
waves through turbulence or sound waves. For runaway walls, there is also a
fraction $\kappa_{\mathrm{wall}}$ of the vacuum energy which goes into kinetic
energy of the wall. This is relevant for gravitational wave generation from
direct bubble collisions. Thus, $\kappa_{\mathrm{fl}}$ is an efficiency
coefficient for the injection of kinetic energy in the fluid, while
$\kappa_{\mathrm{wall}}$ is as an efficiency coefficient for accelerating the
wall.

We have studied, on the one hand, the hydrodynamics of a phase transition
front, for given values of the wall velocity and acceleration, i.e.,
considering these variables as free parameters. Thus, we obtained results for
$\kappa_{\mathrm{wall}}$ and $\kappa_{\mathrm{fl}}$ as functions of the
velocity $v_w$ and the net force $F_{\mathrm{net}}$ acting on the wall. In this
way, the results do not depend on the very involved computation of the wall
dynamics, for which several approximations are generally needed. On the other
hand, we have also studied the wall dynamics, taking into account the
back-reaction of hydrodynamics on the wall motion, and we have discussed on the
calculation of the wall velocity and acceleration as functions of thermodynamic
and friction parameters.

We have computed the efficiency factor $\kappa_{\mathrm{fl}}$  for different
wall symmetries, namely, spherical, cylindrical and planar walls. This
complements the work of Ref.~\cite{lm11}, where we performed a similar analysis
for stationary solutions. Here we considered the runaway case. For planar walls
we obtained analytic results. The result for spherical bubbles is quite similar
to the planar case, which can thus be used as an analytic approximation for the
former. Besides, we provide fits for the spherical case as functions of $v_w$,
$F_{\mathrm{net}}$, and thermodynamic parameters.

For the analysis of the wall dynamics, we considered a phenomenological model
for the friction, which was introduced in Ref.~\cite{ariel13} and depends on
two free parameters. Thus, the model can reproduce the correct value of the
friction force in the NR limit, $F_{\mathrm{fr}}\sim \eta_{NR}\, v_w$ as well
as in the UR limit, $F_{\mathrm{fr}}\sim \eta_{UR}$. In the UR case it is
actually more straightforward, for a given model, to compute the net force
$F_{\mathrm{net}}$. The determination of the friction component of the UR force
is relevant for the use of this phenomenological interpolation, which allows to
calculate the wall velocity away from the UR limit. We have clarified the
decomposition of $F_{\mathrm{net}}$ into driving and friction forces, taking
hydrodynamics effects correctly into account.

Some of the issues discussed in the present paper were previously considered in
Ref.~\cite{ekns10} (for the case of spherical walls). However, a simpler
phenomenological friction was used in that work, which depends on a single
friction coefficient, corresponding to the particular case
$\eta_{NR}=\eta_{UR}$. Moreover, some results, particularly those for the
efficiency factor in the runaway regime, are given in terms of this friction
coefficient (cf. Figs.~10 and 12 in \cite{ekns10}). Therefore, those results
depend on the wall dynamics. As we have seen, the effect of reheating on the
driving force was neglected in Ref.~\cite{ekns10}, which leads to a different
distribution of the released energy. Concrete expressions for the runaway
efficiency factors are given in \cite{ekns10} as functions of $\alpha$ and the
UR detonation limits $ \alpha_0,\kappa^{\mathrm{det}}_{UR}$. In contrast, we
obtained these quantities directly as functions of $\alpha$ and
$F_{\mathrm{net}}$. Thus, we provide  clean results for
$\kappa_{\mathrm{wall}}$ and $\kappa_{\mathrm{fl}}$, which can be used to
compute the production of gravitational waves in a phase transition.

In physical models, strongly first-order phase transitions (i.e., those with a
relatively high value of the order parameter, $\phi>T$), generally have large
amounts of released energy and considerable supercooling. This gives large
values of $\alpha$, as well as high wall velocities or even runaway walls. We
have explored the efficiency factors and the generation of gravitational waves
for such parameter variations.

For detonations and runaway walls, the efficiency factor $\kappa_{\mathrm{fl}}$
decreases with the wall velocity and acceleration, while it increases with the
quantity $\alpha$. However, for given values of the friction parameters, $v_w$
and $F_{\mathrm{net}}$ are increasing functions of $\alpha$, and it turns out
that $\kappa_{\mathrm{fl}}$ decreases with $\alpha$. In contrast,
$\kappa_{\mathrm{wall}}$ increases with $F_{\mathrm{net}}$ and $\alpha$. As a
consequence, the gravitational wave production through fluid motions generally
decreases with $\alpha$, while the production through bubble collisions
generally increases. This does not mean, though, that stronger phase
transitions will be less efficient  in producing gravitational waves through
fluid motions. In our parameter variations we have fixed some quantities which,
for a concrete model, will change as $\alpha$ changes. In Ref.~\cite{lms12} the
electroweak phase transition was considered for several extensions of the
Standard Model (all of which gave steady-state walls). In those cases, stronger
phase transitions gave stronger GW signals from fluid motions. We shall
consider models with even stronger phase transitions in a forthcoming paper
\cite{lm15b}. As we have seen, for parameters which are characteristic of such
strongly first-order electroweak phase transitions, the gravitational waves may
be observed in the planned observatory eLISA.

\section*{Acknowledgements}

This work was supported by Universidad Nacional de Mar del Plata, Argentina,
grant EXA699/14, and by FONCyT grant PICT 2013 No.~2786. The work of L.L. was
supported by a CONICET fellowship.

\appendix{}

\section{Approximations for the fluid efficiency factor} \label{apend}

In Ref.~\cite{lm11} the hydrodynamics was studied for the stationary case, for
spherical, cylindrical and planar walls. Although the fluid profiles are
different in the three cases (corresponding to the spreading of the released
energy in 3, 2, and 1 dimensions, respectively), the total energy in fluid
motions is quite similar, particularly for detonations.  Therefore, a
relatively good approximation for the efficiency factor is to consider a planar
wall, for which one obtains analytic results. One expects the same to hold for
runaway walls.

For planar symmetry, the shape of the rarefaction wave behind the wall is quite
simple. We have constant fluid velocity and enthalpy between the wall and a
point which follows the wall with velocity  $v
_{0}=(v_{-}+c_{-})/(1+v_{-}c_{-})$. Hence, the rarefaction actually begins
behind that point. In the variable $\xi=z/t$, this point lies at a fixed
position $\xi =v_{0}$, while the wall is at $\xi=v_w$.  Between $\xi =c_{-}$
and $\xi =v _{0}$ we have
\begin{equation}
v=\frac{\xi -c_{-}}{1-c_{-}\xi },\quad w=w_{-}\left( \frac{1-v_{-}}{1+v_{-}}
\frac{1+v}{1-v}\right) ^{\frac{1}{2}(c_{-}+\frac{1}{c_{-}})}.
\end{equation}
For the bag EOS, we have $c_-=1/\sqrt{3}$, and the values of $v_-$ and $w_-$
are given by Eqs.~(\ref{vme}-\ref{wmewma}) for detonations and by
Eqs.~(\ref{tmevmerun}) for runaway walls. For this profile, the integral
$I(v_w,v_-)$ in Eqs.~(\ref{kappa}-\ref{kapparun}) is given by \cite{lm11}
\begin{equation}
I=\gamma _{-}^{2}v_{-}^{2}\left( v_{w}-v _{0}\right) +\frac{3\left( 2-
\sqrt{3}\right) ^{\frac{2}{\sqrt{3}}}}{4}\left[ \frac{1-v_{-}}{1+v_{-}}
\right] ^{\frac{2}{\sqrt{3}}}\left[ f\left( v _{0}\right) -f\left(
c_{-}\right) \right] ,  \label{kappapl}
\end{equation}
where $f\left( \xi \right) =\left( \frac{1+\xi }{1-\xi }\right) ^{\frac{2}{
\sqrt{3}}}\left\{ \frac{2}{\sqrt{3}}-1+\left( 1-\xi \right) \left[
2-\,_{2}F_{1}(1,1,\frac{2}{\sqrt{3}}+1,\frac{1+\xi }{2})\right] \right\} $,
and $_{2}F_{1}$ is the hypergeometric function \cite{grads}. The efficiency
factors
$\kappa^{\mathrm{det}}_{\mathrm{pl}},\kappa^{\mathrm{run}}_{\mathrm{pl}}$
for the planar case are obtained by inserting Eq.~(\ref{kappapl}) in
Eq.~(\ref{kappa}) (with $j=0$) or directly in Eq.~(\ref{kapparun}) for the
runaway case. The result is shown in Fig.~\ref{figplesf}, together with that
for spherical walls.

\begin{figure}[bth]
\centering
\epsfysize=7cm \leavevmode \epsfbox{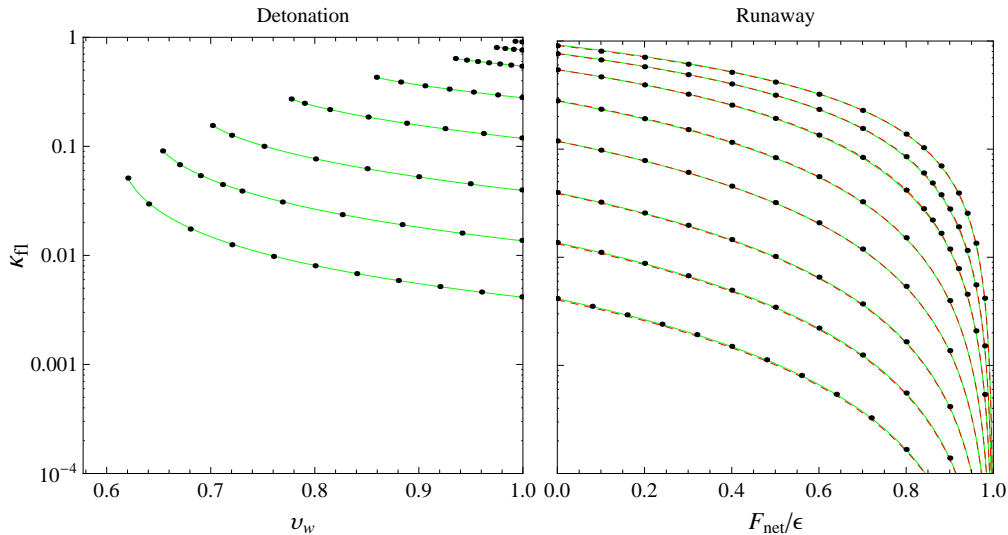}
\caption{Fits for the efficiency factor for  spherical walls, corresponding (from bottom to top) to
$\protect\alpha =0.003,0.01,0.03,0.1,0.3,1,3,10$. The black dots indicate values of the numerical result.
Red dashed lines correspond to the fit of Eq.~(\ref{fit}), and green lines to
the fits of Eqs.~(\ref{fitpl1}-\ref{fitpl2}).}
\label{figfit}
\end{figure}

Since the planar and spherical results are so similar, we may construct a
fit for the spherical case by just approximating the difference between the
two results. Indeed, correcting the planar results with a factor $1.03 + 0.1
\sqrt{v_w - v_J(\alpha)}$ is a good approximation. We thus have
\begin{eqnarray}
\kappa_{\mathrm{fl}}^{\mathrm{det}}&=&\kappa^{\mathrm{det}}_{\mathrm{pl}}(\alpha,v_w)
\left(1.03 + 0.1 \sqrt{v_w - v_J(\alpha)}\right),
\label{fitpl1}
\\
\kappa_{\mathrm{fl}}^{\mathrm{run}}&=&\kappa^{\mathrm{run}}_{\mathrm{pl}}(\alpha,\bar F)
\left(1.03 + 0.1 \sqrt{1 - v_J(\alpha)}\right).
\label{fitpl2}
\end{eqnarray}
In Fig.~\ref{figfit} we compare these fits with the numerical result. In the
whole detonation range, and for $F_{\mathrm{net}}/\varepsilon<0.9$ in the
runaway case, the relative error is smaller than a 3\%.

In the runaway regime  it is easy to find a simple fit (which does not rely
on the analytic formulas of the planar case), since the integral
$I_{1}(v_{-})= I(1,v_-)$ depends on the single parameter $v_-$.  In the
whole range $0<v_{-}<1$, this function is well approximated by the
polynomial $I_{1}(v_{-})\simeq 0.15v_{-}^{2}-0.132v_{-}^{3}+0.065v_{-}^{4}$,
with a relative error which is smaller than a 3\% for $v_->10^{-3}$.
Inserting in Eq.~(\ref{kapparun}), we have
\begin{equation}
\kappa _{\mathrm{fl}}^{\mathrm{run}}\simeq (4/\alpha)(1+3\alpha-3\bar{F}
)(0.15v_{-}^{2}-0.132v_{-}^{3}+0.065v_{-}^{4})\,, \label{fit}
\end{equation}
with $v_-(\alpha,\bar{F})$ given by Eq.~(\ref{tmevmerun}). The result is
shown in the right panel of Fig.~\ref{figfit}.

\end{document}